\documentclass[pre,showpacs]{revtex4}

\def\mathR{\mathbb R}
\def\calD{{\cal D}} 
\def\calK{{\cal K}} 
\def\calS{{\cal S}}
\def\calL{{\cal L}}
\def\calE{{\cal E}}
\def\viz{{\it viz.} }
\def\eg{{\it e.g.} }
\def\tw{}

\def\goesto{\rightarrow}

\usepackage[dvips]{epsfig}
\usepackage{amsmath}
\usepackage{amssymb}
\usepackage{latexsym}
\usepackage{float}

\newcommand{\ulI}{\underline{\rm I}} 
\newcommand{\ulDelta}{\underline \Delta} 
\newcommand{\begineq}[1]{\begin{equation}\label{#1}}
\newcommand{\eqend}{\end{equation}}
\newcommand{\definedas}{\equiv}

\newcommand{\integral}{\int}
\newcommand{\plusorminus}{\pm}
\newcommand{\ie}{{\it i.e.} }

\begin{document}

\title{Singularities, Structures and Scaling in Deformed Elastic
$m$-Sheets}
\author{B.A. DiDonna}
\affiliation{Department of Physics, University of Chicago, Chicago, IL 
60637}
\author{S.C. Venkataramani}
\affiliation{Department of Mathematics, University of Chicago, Chicago, IL 
60637}
\author{T.A. Witten}
\affiliation{Department of Physics, University of Chicago, Chicago, IL 
60637}
\author{E.M. Kramer}
\affiliation{Department of Natural Sciences and Mathematics, 
Simon's Rock College, Great Barrington, MA 01230}

\begin{abstract}{
The crumpling of a thin sheet can be understood as the condensation of
elastic energy into a network of ridges which meet in
vertices. Elastic energy condensation should occur in response to 
compressive strain in elastic objects of any dimension
greater than $1$. We
study elastic energy condensation  numerically
in $2$-dimensional elastic sheets
embedded in spatial dimensions $3$ or $4$ and 
$3$-dimensional elastic sheets embedded in spatial dimensions $4$ and
higher. We represent a sheet as a lattice of nodes with an
appropriate energy functional to impart stretching and bending
rigidity. Minimum energy configurations are found 
for several different sets of boundary conditions.
We observe two distinct behaviors of local energy density
fall-off away from singular points, which we identify as cone scaling or
ridge scaling.
Using this analysis we demonstrate that there are marked
differences in the forms of energy condensation depending
on the embedding dimension. 
}
\end{abstract}









\maketitle


\section{Introduction}
\label{sec:intro}

In the last several years, there has been a marked interest in 
the nature of crumpling~\cite{nelson, Seung.Nelson,
our.stuff, Pomeau, Pomeau2, maha,boudaoud,origami,
chaieb.cone,maha.cone,chaieb.crescent}.  Field theories have been 
formulated for the crumpling transition~\cite{nelson},
quantitative laws have been educed for the
energy scaling of crumpled sheets~\cite{our.stuff,Pomeau,maha},
and dynamics of the crumpling process have been simulated and 
measured~\cite{boudaoud}. In this paper we treat crumpling as an
example of energy condensation.

The crumpling of a thin elastic sheet
can be viewed as the  condensation of elastic energy  
onto a network of point vertices and folding ridges.
These structures spontaneously emerge, for example,
when a thin sheet of 
thickness $h$ and spatial extent $L \gg h$ is confined within a ball of 
diameter $X < L$. For $X \le L/2$ the 
important length scales become $h$ and $X$. The 
elastic energy scaling of vertices and ridges are well 
understood~\cite{our.stuff,maha}.
In the limit $h/X \rightarrow 0$, the elastic energy 
is believed to condense into a 
vanishingly small area around the ridges and vertices.

There is a significant body of physics literature on energy
condensation, because it is a pervasive feature of condensed
matter. This behavior is seen in many systems including type-two
superconductors \cite{DeGennes.superconductor}, strongly turbulent
flow \cite{intermittency.review} as well as in mechanical
\cite{fracture} and electrical \cite{dielectric.breakdown} material
failure.  Analogous condensation also occurs in particle-confining
gauge field theories \cite{asymptotic.freedom}. 
In a mathematical context, such condensation often
arises in singular perturbations of non-convex variational problems
\cite{variational.review,sternberg}. A few examples of such problems are the
gradient theory of phase transitions \cite{phase_transitions}, wherein
the bulk of the energy is condensed into a small neighborhood of the
interface between the two phases; Ginzburg-Landau vortices
\cite{Ginzburg.Landau} which, among other things, describe type-two
superconductors; and solid-solid phase transitions in crystalline
materials (martensitic phases) \cite{james,variational.review}. 

One distinctive aspect of the energy condensation in crumpling 
is the interesting dependence of the total energy 
scaling on boundary conditions.  A survey of the elastic energy 
scaling with thickness $h$ for a given material with $2$-dimensional
strain modulus $\mu$
illustrates this point.
For elastic sheets that are forced so that they
form a single conical vertex or
``$d$-cone'' \cite{Pomeau}, the only curvature singularity in the
$h \rightarrow 0$ limit is at the vertex of the cone
\cite{Pomeau,chaieb.cone,maha.cone}. The total energy of the
sheet scales as $\mu h^2 \log(X/h)$ \cite{maha.cone} in this
situation. When the boundary conditions are such that there are many 
vertices and ridges (\eg confinement), 
the elastic energy is concentrated on the ridges.
For confined sheets, the typical ridge length is on the order of the 
confining diameter $X$. It has been argued that 
ridges with length $X$ have a
characteristic total elastic energy which 
scales as $\mu h^{5/3}X^{1/3}$ \cite{Alex,science.paper}, and 
that the total energy of the
system scales 
with the same exponent. A final example is the delamination
and blistering of thin films, which is
described by the same energy functional as the crumpled sheet but
with different boundary 
conditions~\cite{GO,AG,KJ,Audoly,DKMO,ADM,JS1,JS2,BCDM}.
In this circumstance, the sheet develops a
self-similar network of folding lines, whose lengths grow smaller as
we approach the boundary \cite{JS2,BCDM}, and the
total energy of the sheet scales as $\mu h$ (with a finite fraction of the
energy concentrating in a narrow layer near the boundary, of a width
that also scales as $h$ \cite{BCDM}).

Thus, by varying the
boundary conditions, the same energy functional can lead to
significantly different forms of energy condensation, with different
energy scalings and different types of energy bearing structures. This
behavior is contrary to the widely held view that singularities are
``local'' phenomena.
Our goal is to study this phenomenon, with a hope
of understanding the factors that determine the nature of energy
condensation in general systems. In this paper, we study elastic
energy condensation in spatial dimensions above three.  Our motivation
is to understand how the scaling behavior of crumpled sheets and the
topology of energy condensation networks generalize for
$m$-dimensional elastic manifolds in $d$-dimensional space.
To this end, our numerical study explores energy condensation in 
$2$-sheets in $3$ or $4$ dimensions
and $3$-sheets in dimensions $4$ -- $6$, subject to boundary 
conditions which are akin to confinement.

Our previous work\cite{eric.math, eric} showed that the notion of an
elastic membrane extends naturally to different dimensions.  Such
membranes have an energy cost for ``stretching" deformations that
change distances between points in the $m$-dimensional manifold and
have an additional cost for bending into the embedding space.  When
these costs are isotropic, the material properties may be expressed in
terms of a stretching modulus, a bending stiffness, and a``Poisson's
ratio" of order unity.  As in $2$-dimensional manifolds, the ratio of
bending stiffness to stretching modulus yields a characteristic
length.  Indeed, if the manifold is a thin sheet of isotropic
$d$-dimensional material, the thickness $h$ of the sheet is a
numerical multiple of the square root of the modulus ratio that may be
readily calculated\cite{eric.math}.  

Our previous paper~\cite{shankar.witten}
identified two regimes of dimensionality with
qualitatively different response to spatial confinement. The 
authors 
considered an elastic $m$-dimensional ball
of diameter $L$ geometrically confined within 
$d$-spheres of diameter less than $L/2$.
When the embedding
dimension $d$ is twice the manifold dimension $m$ or more, the state of
lowest energy is one of non-singular curvature,
with stretching elastic energy indefinitely smaller than bending energy.
For the complementary cases where $d$ is
smaller than $2m$, the deformation is qualitatively different.  Such
manifolds cannot be geometrically
confined in a sphere of diameter smaller than $L/2$
without stretching or singular curvature. 
In ordinary $2$-sheets ($m =2$) in $3$ dimensions
energy condenses in order to reduce the stretching energy of 
spatial confinement.  The
degree of energy condensation depends on the stretching moduli through
the thickness $h$ defined above. In $3$-sheets, singularities or stretching
are required in $4$ or $5$ embedding dimensions.
Previous work~\cite{eric} confirmed
that for $3$-sheets confined in $4$ dimensions, energy condenses into a
network of line-like vertices and planar ridges. We seek to understand
how the degree of energy condensation associated with confinement
changes with increasing spatial dimension. We expect that less energy
will be required to confine a $3$-sheet within a $5$-dimensional sphere
than within a $4$-dimensional sphere, but we do not know {\em a priori}
how the form of energy condensation will differ between these two cases.

We begin our study
by giving a brief review of elastic theory in Section~\ref{sec:theory}. 
Then, Section~\ref{sec:structures} quantifies our
definitions of ``folding lines'' and ``vertices'' within a framework of 
isometric embeddings, and in Section~\ref{sec:isometric}
we propose a rule for the topological 
dimensionality of vertices in energy condensation networks.
In Section~\ref{sec:scaling} 
we present analytical estimates for the degree of
energy condensation in the crumpled state. Building on existing
knowledge, we  
make predictions for the scaling of energy
density with distance away from the regions of greatest elastic energy.
We identify two distinct forms of energy scaling, which we call ridge
scaling and cone scaling (the names are
based on the geometry these scalings correspond to in
ordinary crumpling of $2$-sheet in $3$ dimensions).
Section~\ref{sec:numerics} 
describes how we represent elastic manifolds
numerically.  

Then we present our numerical findings. 
We begin in Section~\ref{sec:confined} with simulations of sheets confined
by shrinking hard wall potentials.
In this qualitative study the embedding dimension seems to affect the
crumpled structure significantly.  The condensation of energy appears to
become progressively weaker as the embedding dimension is increased,
culminating in no condensation when $d$ reaches $2m$. Numerical 
difficulties prevented any significant quantitative analysis of the
geometrical confinement data. The need for better data motivates the 
simpler systems we simulated next.

Section~\ref{sec:disclin} describes our studies of $m$-sheets
with two disclinations.  
Disclination are made by removing wedge shaped sectors from the sheet
and then joining the edges of each wedge.  The essential
feature of a disclination is that it induces the sheet to form a cone,
with lines of null curvature converging at a vertex.  
It has been shown that when {\it two}
disclinations are introduced into a $2$-sheet in $3$ dimensions, 
the elastic energy of deformation between the disclinations condenses
along a ridge joining the two vertices~\cite{our.stuff}.  
These ridges appear completely similar to those in geometrically
confined sheets and exhibit the same energy 
scaling~\cite{Alex}. 
In our present study, simulated $2$-sheets in $3$ dimensions
formed the familiar ridges, but $2$-sheets with the same boundary
conditions in $4$-dimensional space had much lower total elastic
energies and very different energy distributions.
Similarly, $3$-sheets in $4$ spatial dimensions formed ridges
closely analogous to those seen in $2$-sheets, but for
$3$-sheets in $5$ dimensions no ridges were evident. 
Also, non-parallel disclination lines in
$3$-sheets appear to generate further disclination-like lines
in $4$ spatial dimensions but not in $5$.

Next, in Section~\ref{sec:torus} 
we detail our simulations of $3$-tori allowed to
relax in $d$ dimensions. The benefit of this geometry is that we 
expect it to cause energy condensation without the need to introduce
disclinations.
Observing that a $2$-torus cannot be smoothly and
isometrically embedded in a space of 
dimensionality less than $4$, we expect an elastic sheet with the
connectivity of an $m$-torus embedded in a space of dimension 
$d < 2m$ will relax to a configuration with 
regions of non-zero strain (condensed into a network of ridges).
We found that a $3$-torus in $d=4$ 
spontaneously forms a network of planar ridges which intersect in 
vertex
singularities similar to those in the geometrically 
confined sheets. 
In $d=5$, the $3$-torus forms a point-like
vertex network with no observable
ridges. The energy scaling and presence or absence of ridges mirrored
the behavior of sheets with
disclinations in Section~\ref{sec:disclin}. 
The complexity of the crumpling network decreases with increasing
embedding dimension, 
with spontaneous symmetry breaking evident for $d=5$. As expected, the
elastic energy distribution is homogeneous for $d \ge 6$. 

Finally, in Section~\ref{sec:singlefold} we present the results of
simulations of a ``bow configuration'', in which the center points of
opposite faces of a $3$-cube were attached and the cube was embedded
in $4$ or $5$ spatial dimensions. With proper manipulation of
initial conditions, the cube embedded in $5$ dimensions forms a
single, point-like vertex at its center. The energy density scaling away
from this singularity agrees with predictions for a novel kind of
elastic structure which is a generalization of a simple cone. By
contrast, the cube embedded in $4$ dimensions forms a set of line-like
vertices and planar ridges
that are well modeled by our present understanding of
$3$-dimensional crumpling.

We conclude by discussing the observed energy scaling properties of
crumpled elastic sheets. We have developed a means 
to identify the presence of ridges in $m$-sheets 
based solely on their spatial elastic energy distribution. 
Using the analysis of 
energy distributions, we demonstrate that folding lines 
in greater than $m+1$ dimensions have
different energy and thickness scaling properties than in $m+1$,
but ridges in $m+1$ seem to have the same scaling regardless of $m$.
We found that ridge scaling dominates the crumpling of $m$-sheets in 
$m+1$ dimensions, while cone scaling
was the only form of scaling witnessed
in dimensions greater than $m+1$.
Differences
in the morphology of higher dimensional folding lines is discussed
briefly. The local structure of 
these folds is very different from that of the familiar ridges found
in $2$-sheets in $3$-dimensional space.
We also note that our simulational
findings strongly support the new rule for the topology of elastic energy 
bearing structures in higher dimensions which is presented in 
section~\ref{sec:isometric}. We end with a brief discussion of the
mathematical questions raised by the non-local character of energy
scaling in crumpled sheets.


\section{Elastic $m$-sheets in $d$-space}
\label{sec:theory}

In this section, we review the elastic theory of $m$-sheets in
$d$-space as it is presented in Ref.~\cite{eric.math}. In analogy with
the elastic 2-sheets of everyday experience, an $m$-sheet in $d$
dimensional space is an elastically isotropic $d$-dimensional solid
which has a spatial extent of order $L$ in $m$ independent directions
and $h \ll L$ in the remaining $d-m$ directions. Specifically, our
$m$-sheet is given by $\calS \times B^{d-m}_h \subset {\mathbb
R}^d$, where $\calS \subset {\mathbb R}^m$ is a set that has a
typical linear size $L$ in all directions, and $B^{d-m}_h$ is a $d-m$
dimensional ball of diameter $h$. 

We are considering embeddings of the $m$-sheet in a $d$-dimensional
target space. We first consider the lowest energy embedding in a
sufficiently large $d$-dimensional space, say all of ${\mathbb R}^d$, so
that the sheet is not distorted in the embedding. We assume that the
undistorted sheet has no intrinsic strains, curvatures or
torsions(twists). Since there are no curvatures or torsions, picking a
orthogonal basis of $d-m$ vectors for the thin directions at one point
on the sheet, and then parallel transporting these vectors to every
point on the sheet gives an orthonormal set of basis vectors for every
point of the sheet. We can therefore describe the geometry of the
undistorted sheet, which is a $d$-dimensional object, by the
$m$-dimensional center surface $\calS$ which gives the geometry in
the long directions, and the orthonormal basis vectors for the thin
directions, that describes the geometry in the thin directions. These
basis vectors for the thin directions give a normal frame field to the
embedding of the center surface, since they are all orthogonal to each
of the long directions in the sheet. Further, since the basis vectors
at different points are related by parallel transport in ${\mathbf
R}^d$, the normal frame field is torsion-free.

For small distortions, we can continue to describe the embedding of
the $m$-sheet by giving the embedding of the center surface $\calS$
and by specifying the normal frame field \cite{eric.math}. The
rotational invariance in the thin directions implies that the torsion
of the sheet in the embedding cannot couple to the geometry of the
center surface. Since the sheet has no intrinsic torsion, if there are
no applied applied torsional forces, the normal frame has to remain
torsion free. The torsion degrees of freedom therefore drop out of the
energetic considerations that will determine the geometry of the sheet
\cite{eric.math}. Therefore, we can leave out the thin directions and
determine the energy of the embedding through an effective Lagrangian
that only depends on the long directions, {\em i.e.}, the geometry of
center surface of the sheet $\calS$, as embedded in the $d$-space
\cite{eric.math}. This approach puts powerful tools of differential
geometry at our disposal.  Numerically, this treatment greatly
increases the efficiency of our simulations by decreasing the
dimension and required grid resolution of our lattice.  In the limit
$h/L \ll 1$ and for relatively small elastic distortions of the
material, this description is highly accurate.

We use Cartesian coordinates in the center surface, which can be
viewed as the set $\calS \subset {\mathbb R}^m$. We refer to these
coordinates as the material coordinates, and quantities referred to the
material co-ordinates will be denoted by Roman subscripts \eg
$i,j,k,l$.  The configuration of the sheet is given by a vector valued
functions $\vec{r}(x_i)$ with values in the $d$-dimensional target
space.  We also denote the $d-m$ normal vectors in a choice for an
orthonormal, torsion-free frame by $\vec{n}^{(\alpha)}$, with a Greek
superscript that takes values $1,2,\ldots,d-m$. Such a choice exists
by our previous considerations.

The strain energy density $\calL_s$ due to the distortions within
the $m$-sheet is given by the conventional expression \cite{Lame.ref}
in terms of the Lam\'e coefficients $\lambda$ and $\mu$
\begineq{lame.energy}
\calL_s = \mu \gamma_{ij}^2 + \frac{\lambda}{2}\gamma_{ii}^2,
\end{equation} 
where $\gamma_{ij}$ is the strain tensor, defined by
$$
\gamma_{ij} = \frac{1}{2} \left(\frac{\partial \vec{r}}{\partial x_i}
\cdot \frac{\partial \vec{r}}{\partial x_j}-\delta_{ij} \right).
$$
The strain tensor quantifies 
the deviation of the metric tensor of the embedded
sheet from it's intrinsic metric tensor. 
Here and henceforth,
repeated indices (both Greek and Roman) are summed over all the range
of their allowed values.  

The non-zero 
thickness of the $m$-sheet 
leads to an energy cost for
distortions of the center surface $\calS$ in a normal direction,
{\em i.e.}, bending distortions. A measure of the bending of the
manifold at any point is the extrinsic curvature tensor
$\vec{\kappa}_{ij}(x_i)$, which is the projection of the second
derivatives $\partial_i \partial_j \vec{r}$ into the normal frame. The
component of the extrinsic curvature in the normal direction
$\vec{n}^{(\alpha)}$ is given by
\begineq{curvature.as.derivative}
\kappa_{ij}^{(\alpha)} = \frac{\partial^2 \vec{r}}{\partial x_i 
\partial x_j} \cdot \vec{n}^{(\alpha)}.
\end{equation}
As shown in Ref.~\cite{eric.math}, if the strains are small and the
curvatures are small compared to $1/h$, the energy density of the
bending distortions $\calL_b$ is given by
\begineq{curvature.energy}
\calL_b = B \left[ \kappa_{ij}^{(\alpha)} \kappa_{ij}^{(\alpha)} +
\frac{\lambda}{2\mu} \kappa_{ii}^{(\alpha)} \kappa_{jj}^{(\alpha)}
\right].
\end{equation}
The bending modulus $B$ in the above
equation is determined by the Lam\'e coefficient $\mu$ and the
thickness of the sheet $h$ through the relation~\cite{eric.math}
$$
B = \mu h^2/\eta (m,d), 
$$ 
where $\eta(m,d)$ is given by
\begineq{eta.def}
\eta(m,d) = \frac{d-m}{S_{d-m}} \times \left\{
\begin{array}{lr}
\frac{2}{3} & d-m=1 \\
\frac{\pi}{4} & d-m=2 \\
\frac{1}{d-m+2} \beta(3/2,d-m-2) S_{d-m-1} & d-m>2
\end{array} \right. ,
\end{equation}
where $S_a = 2 \pi^{a/2}/\Gamma (a/2)$ is the area of a unit sphere in
$a$ dimensions and $\beta(a,b) = \Gamma(a) \Gamma(b) / \Gamma(a+b)$ is the
beta function.
For $m=3$ and $d=4,5,6$, 
$\eta(m,d)=3,4$ and $5$ respectively.

For studying the geometrical confinement of an elastic $m$-sheet, the
confining forces are assumed to be derived from a potential
$V_{c}(\vec{r})$ in the embedding space. The energy of the $m$-sheet
is the sum of the bending energy, the strain energy and the energy due
to the spatially confining potential. Therefore, the total energy is
given in terms of the geometry of the center surface $\calS$ by
\begineq{confinement}
{\cal E}  = \mu \integral_\calS d^m x \left[ \frac{h^2}{\eta} \left( 
\kappa_{ij}^{(\alpha)} \kappa_{ij}^{(\alpha)}
+ \frac{\lambda}{2 \mu} \kappa_{ii}^{(\alpha)} \kappa_{jj}^{(\alpha)}
\right) +
\left (\gamma_{ij}^2 + \frac{\lambda}{2\mu} \gamma_{ii}^2 \right ) + 
\frac{V_{c}(\vec{r}(x_i))}{\mu} \right],
\end{equation}
where $\eta = \eta(m,d)$ as defined in Eq.~\ref{eta.def}.

The configuration $\vec{r}(x_i)$ of the sheet in the embedding space
is obtained by minimizing the energy ${\cal E}$ over the set of all
allowed configurations. A (local) minimum energy configuration is
obtained by requiring that the variation $\delta {\cal E}$ should
vanish to the first order for an arbitrary (small) variation $\delta
\vec{r}$ of the configuration. Since the energy density contains terms
in $\kappa_{ij}$, that involve the second derivatives of the function
$\vec{r}(x_i)$, the Euler-Lagrange equations for the minimization
problem are a system of fourth order, nonlinear elliptic equations on
the domain $\calS$. Very little is known about the rigorous
analysis of such equations. Therefore, we will study the geometrical
confinement problem numerically, by approximating the the integral in
Eq.~\ref{confinement} by a sum over a grid, and minimizing the
resulting energy by a conjugate gradient method \cite{conj_grad}, as we
outline below.

Our goal is to study the scaling behavior of the structures on which
the energy concentrates as the thickness $h \rightarrow 0$. The
variational derivative of the potential term is given by
$$
\frac{\delta}{\delta \vec{r}} \int_\calS d^m x V_c(\vec{r}(x_i)) =
\nabla_{\vec{r}} V_c(\vec{r}(x_i)).
$$
This term leads to a strongly non-linear coupling between the
configuration of the minimizer and the stresses and the bending
moments in the sheet. Consequently, the conditions for mechanical
equilibrium are now ``global'' and the stresses and bending moments
determined by the local strains and curvatures should balance a term
that depends on the global geometry of the configuration. In addition
to complicating the analysis, this introduces length scales besides
the thickness $h$ into the problem. This in turn can lead to the lack
of simple scaling behavior at equilibrium for the structures in
geometrically confined sheets. Note however, that this is not the case
for confinement in a hard wall potential
$$
V_c(\vec{r}) = \left\{ \begin{array}{cc} V_0 & \mbox{ for } \vec{r} \in
\Omega \\ + \infty & \mbox{ otherwise } \end{array} \right.
$$
where $\Omega$ is a given set in ${\mathbb R}^d$. The configuration of
the minimizer is now restricted to be inside $\Omega$ and the gradient of
$V_c$ is zero here, so that there is no coupling between the the
configuration of the minimizer and the stresses and the bending
moments in the sheet, in the parts of the sheet that are in the
interior of $\Omega$.

One way to get around this problem is to study the configurations
where the energy concentration is due to the boundary conditions
imposed on the sheet, and not due to an external potential. If the
imposed boundary conditions do not introduce any new length scales, we
would then expect to see structures and scalings that are generic,
{\em i.e.}, independent of the precise form of the imposed boundary
conditions. This is analogous to the minimal ridge \cite{Alex} that is
obtained by imposing boundary conditions on a 2-sheet. Although the
minimal ridge is obtained with a specific boundary condition, the
scaling behaviors of the ridge are generic and are seen with a variety
of boundary conditions.

In this study we first determine the generic structures and scalings
that we expect to see for an $m$-sheet in $d$-dimensional space. We
also numerically verify our predictions for these scalings by the
configuration of an embedded $m$-sheet with a variety of boundary
conditions -- sheets with disclinations, sheets with a toroidal global
connectivity, and sheets in a ``bow'' configuration. In all these
cases, the elastic energy is given by
\begineq{eq:energy} 
{\cal E} = \mu \integral_\calS d^m x \left[ \frac{h^2}{\eta} \left(
\kappa_{ij}^{(\alpha)} \kappa_{ij}^{(\alpha)} + \frac{\lambda}{2\mu}
\kappa_{ii}^{(\alpha)} \kappa_{jj}^{(\alpha)} \right) + \left
(\gamma_{ij}^2 + \frac{\lambda}{2\mu} \gamma_{ii}^2 \right ) \right].
\end{equation} 
However, the domain of integration $\calS$ is no longer a subset of
${\mathbb R}^m$. It is a domain with singularities in the case of
sheets with disclinations or in the ``bow'' configuration, or a set
whose global topology is different from ${\mathbb R}^m$, in the case
of the sheets with toroidal connectivity. Note that this energy
functional is also applicable to the confinement in a hard wall
potential, since, without loss of generality, we can set $V_0 = 0$ for
$\vec{r}$ in $\Omega$, and impose the constraint of the hard wall
potential through the conditions $\vec{r}(x_i) \in \Omega$ for all
$x_i \in \calS$. Consequently, the energy is still given by
Eq.~\ref{eq:energy}, and the energy condensation is due to the
additional constraints that are imposed, that are analogous to the
boundary conditions considered above.

We can rewrite the energy using the in-plane stresses $\sigma_{ij}$
and the bending moments $M^{(\alpha)}_{ij}$ that are conjugate to the
strains $\gamma_{ij}$ and the curvatures $\kappa^{(\alpha)}_{ij}$
respectively. The conjugate fields are given by the variational
derivatives
$$
\sigma_{ij} = \frac{\delta {\cal E}}{\delta {\gamma_{ij}}} = 2 \mu
\gamma_{ij} + \lambda \delta_{ij} \gamma_{kk},
$$ 
and
$$
M^{(\alpha)}_{ij} = \frac{\delta {\cal E}}{\delta
{\kappa^{(\alpha)}_{ij}}} = \frac{h^2}{\eta} \left(2 \mu
\kappa^{(\alpha)}_{ij} + \lambda \delta_{ij} \kappa^{(\alpha)}_{kk}\right),
$$
where we have taken the variational derivatives as though the fields
$\gamma_{ij}$ and $\kappa^{(\alpha)}_{ij}$ are independent. The energy
can now be written as
$$
{\cal E} = \frac{1}{2} \int_\calS \left(M^{(\alpha)}_{ij}
\kappa^{(\alpha)}_{ij} + \sigma_{ij} \gamma_{ij} \right) d^m x.
$$ 
Although the energy functional ${\cal E}$ does not explicitly couple
the strains in the manifolds to the curvatures (See
Eq.~\ref{eq:energy}), they are related by geometric constraints
since they are both defined by derivatives of the embedding
$\vec{r}(x_i)$. 
Since the $m$-sheet is intrinsically flat, the Riemann curvature
tensor for the embedding of the center surface can be expressed in
terms of the {\it extrinsic} curvature $\kappa^{(\alpha)}_{ij}$ by the
Gauss Equation \cite{diff_geom},
$$
R_{ijkl}[\kappa] = \kappa_{ik}^{(\alpha)}\kappa_{jl}^{(\alpha)} -
\kappa_{il}^{(\alpha)}\kappa_{jk}^{(\alpha)}.
$$
However, the Riemann curvature is intrinsic to the geometry of the
center surface, and can be written in terms of the strains as 
$$
R_{ijkl}[\gamma] = -\gamma_{ik,jl} + \gamma_{il,jk}-\gamma_{jl,ik} +
\gamma_{jk,il} + O(\gamma^2).
$$
Consequently, the curvatures $\kappa^{(\alpha)}_{ij}$ and the strains
$\gamma_{ij}$ are constrained in order that $R_{ijkl}[\kappa] =
R_{ijkl}[\gamma]$. 

From the symmetries of the Riemann tensor, it has $m(m-1)(m^2-m+2)/8$
independent components. However, since it can be written purely
as a function of the strain $\gamma_{ij}$, it can only have as many 
independent degrees of freedom as the strain itself. As noted in
Ref.~\cite{eric.math}, the strain tensor is symmetric, and further it
satisfies $m$ additional conditions from the balance of in-plane
stresses. Consequently, the strain has $m(m-1)/2$ independent
components, and this yields $m(m-1)/2$ independent
constraints on the extrinsic curvatures.

For $m = 2$, {\em i.e.}, for 2-sheets, there is one constraint, and
this is most economically expressed through the Gaussian curvature of
the sheet \cite{eric.math}. In terms of the extrinsic curvatures, the
Gaussian curvature $G$ is given by
$$
G[\kappa] = \kappa^{(\alpha)}_{11}\kappa^{(\alpha)}_{22} -
\kappa^{(\alpha)}_{12}\kappa^{(\alpha)}_{12},
$$
and in terms of the strains, the Gaussian curvature is given by
$$
G[\gamma] = -\gamma_{11,22} + 2 \gamma_{12,12} - \gamma_{22,11}
+ O(\gamma^2).
$$
We can impose the constraint $G[\kappa] = G[\gamma]$ through a
Lagrange multiplier $\chi$, so that the augmented energy functional is
now given by \cite{eric.math}
$$
{\cal E}_{\chi} = \int_\calS d^m x \left[\frac{1}{2}
\left(M^{(\alpha)}_{ij} \kappa^{(\alpha)}_{ij} + \sigma_{ij}
\gamma_{ij} \right) + \chi (G[\gamma] - G[\kappa])\right].
$$ 
Taking the variations with respect to $\gamma_{ij}$, $\chi$ and
$\kappa_{ij}^{(\alpha)}$ give
\begin{eqnarray}
\sigma_{ij} & = & \delta_{ij} \nabla^2 \chi - \partial_i \partial_j
\chi \Rightarrow \partial_i \sigma_{ij} = 0, 
\nonumber 
\\ 
G[\gamma] & = & G[\kappa], 
\nonumber 
\\ 
\mbox{and} \quad \partial_i \partial_j
M^{(\alpha)}_{ij} & = &\sigma_{ij} \kappa^{(\alpha)}_{ij},
\nonumber
\end{eqnarray}
which are respectively, the balance of the in-plane stresses, the
Geometric (or First) von Karman equation and the Force (or Second) von
Karman equation \cite{vK,eric.math}. The first equation also shows
that the Lagrange multiplier $\chi$ is the scalar stress function of
Airy \cite{Airy}.

In this work, we will mainly focus on the case $m > 2$. For $m > 2$,
an economical way to impose the geometric constraint relating the
extrinsic curvatures to the strains is through the Einstein tensor
\cite{eric.math}
$$
G_{ij} = R_{ikjk} - \frac{1}{2} \delta_{ij} R_{lklk},
$$
which has $m(m-1)/2$ independent components since it is symmetric
and satisfies the contracted Bianchi identity \cite{waldbook}
$$
\partial_i G_{ij} = 0.
$$
In terms of the extrinsic curvature, 
\begineq{gvk_curv}
G_{ij}[\kappa] = \kappa^{(\alpha)}_{ij} \kappa^{(\alpha)}_{kk} -
\kappa^{(\alpha)}_{ik} \kappa^{(\alpha)}_{jk} - \frac{1}{2}
\delta_{ij} \left[ \kappa^{(\alpha)}_{ll} \kappa^{(\alpha)}_{kk} -
\kappa^{(\alpha)}_{lk} \kappa^{(\alpha)}_{lk} \right] ,
\end{equation}
and to the first order in the strains 
\begineq{gvk_strain}
G_{ij}[\gamma] =
-\gamma_{ij,kk}+\gamma_{ik,jk}-\gamma_{kk,ij}+\gamma_{kj,ik} +
\delta_{ij}\left[\gamma_{ll,kk} - \gamma_{lk,lk} \right].
\end{equation}
As in the case $m = 2$, the constraint $G_{ij}[\kappa] =
G_{ij}[\gamma]$ is incorporated through a tensor Lagrange multiplier
$\chi_{ij}$. The augmented energy functional is given by
$$
{\cal E}_{\chi} = \int_\calS d^m x \left[\frac{1}{2}
\left(M^{(\alpha)}_{ij} \kappa^{(\alpha)}_{ij} + \sigma_{ij}
\gamma_{ij} \right) + \chi_{ij} (G_{ij}[\gamma] - G_{ij}[\kappa])\right].
$$ 
Taking the variations with respect to $\gamma_{ij}$, $\chi_{ij}$ and
$\kappa_{ij}^{(\alpha)}$ give the balance of in-plane stresses
$$
\partial_i \sigma_{ij} = 0, 
$$
the Geometric von Karman equation
\begin{equation}
G_{ij}[\gamma]  =  G_{ij}[\kappa],
\label{eq:GvK}
\end{equation}
and the Force von-Karman equation
\begin{equation}
\partial_i \partial_j M^{(\alpha)}_{ij}  = \sigma_{ij}
\kappa_{ij}^{(\alpha)},
\label{eq:FvK}
\end{equation}
respectively \cite{eric.math}.  In the case $m=3$, the Lagrange
multiplier $\chi_{ij}$ is the Maxwell stress function~\cite{Maxwell}.

\section{Structures in elastic $m$-sheets} 
\label{sec:structures}

We will now investigate the minimum energy configurations of the sheet
with external forcing. As we discussed earlier, the sheet can be
forced either by an external potential $V_c(\vec{r})$ (See
Eq.~\ref{confinement}) or by restricitng the set of admissible
configurations by appropriate boundary conditions
(Eq.~\ref{eq:energy}). Since confinement by a hard wall potential of
radius $r_0$ is also given by the energy functional in
Eq.~\ref{eq:energy} where the admissibility condition is that
$\|\vec{r}(x)\| \leq r_0$ for all $x \in \calS$, we will restrict
our attention to the energy functional ${\cal E}$ in
Eq.~\ref{eq:energy}.

From Eq.~\ref{eq:energy}, we see that the only length scales in the
energy functional are the thickness $h$ and the length scale $L$ that
is associated with the center surface $\calS$. 
Since the effective bending modulus $\mu h^2/\eta$ goes to zero as $h
\rightarrow 0$, except in the vicinity of regions with large
curvature, the large scale (O($L$)) behavior of crumpled sheets
should be determined almost entirely by the stretching energy
functional
$$
{\cal E}_s = \mu \int_\calS \left (\gamma_{ij}^2 + c_0
\gamma_{ii}^2 \right) d^m x,
$$
which penalizes the deviation of the configuration from an
isometry. 
Indeed, crumpled $2$-sheets in $3$ dimensions can be described as a
set of nearly isometric regions bounded by areas of large curvatures
that include  vertices and  boundary layers around folds. 
Since the curvature in these regions
is large, the bending energy in this region will continue to remain
relevant as $h \rightarrow 0$. As $h \rightarrow 0$, the width of, and
the strain in, the boundary layer around folds goes to zero,
and the only non-isometric regions are the vertices.

For the remainder of this paper we assume that in any dimension,
the minimum energy configurations of
crumpled $m$-sheets converge in the $h \rightarrow 0$ limit to 
configurations which are locally isometric and have smooth, well
defined curvature almost everywhere. In this view, the regions of
elastic energy concentration in the $m$-sheets converge in the $h
\rightarrow 0$ limit to a
{\em defect set} in the manifold which is not locally smooth and
isometric, and this defect set is as small as
possible relative to the boundary conditions imposed on the sheet.
The limiting procedure which connects the defect set to the energy
concentration regions is elaborated upon in Sec.~\ref{sec:isometric}.

These assumptions
give us descriptive tools to classify the elastic energy structures in
higher dimensional crumpled sheets in terms of well defined concepts of
isometry. More importantly, the identification of crumpling with
isometric embedding will allow us to make predictions for the
dimensionality of energy condensation regions in $m$-sheets for $m >
2$ based on geometric results on isometric immersions. We
will present these arguments in Sec.~\ref{sec:isometric} and the
numerical studies reported in
Sections~\ref{sec:disclin}~-~\ref{sec:singlefold} appear to support
these predictions.

\begin{figure}[tbp!]

{

\center
\epsfig{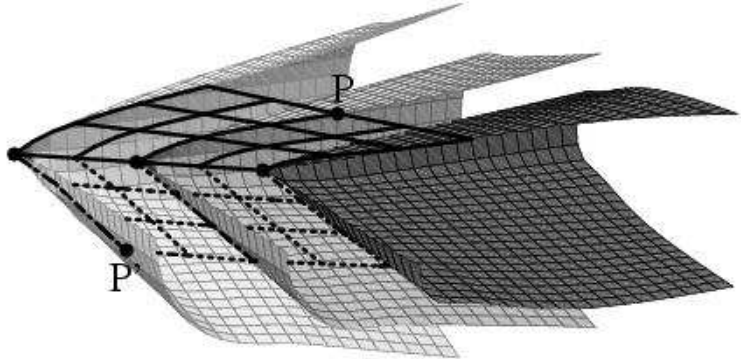}

}

\caption{
Flat subspaces $R$ in a crumpled sheet.  The figure
shows sections of three parallel planes of a thin 3-cube confined in
four dimensions by boundary conditions discussed in
Section~\ref{sec:singlefold}. 
This sheet is nearly 
isometric over most of its volume, as anticipated in
Section~\ref{sec:isometric}.  The 
arguments of this section suggest that such sheets should have a nearly-flat 
$2$-dimensional subsheet $R$ through any point $p$.  The $R$ for the 
indicated point $p$ is shown as a solid grid.  The nearly-flat subsheet $R'$ 
for a different point $p'$ is shown as a dashed grid.  The adjacent 
boundaries of $R$ and $R'$ meet in a nearly-straight, $1$-dimensional 
region.  We identify this region as a vertex.
}
\label{fig:butterfly}
\end{figure}

\subsection{Dimensionality of Defects} \label{sec:isometric}

In this section we define a certain type of singularity called a vertex 
that must exist in confined $m$-sheets.  We then argue that a vertex must 
have a dimensionality of at least $2m - d - 1$.  
In previous work~\cite{shankar.witten}
we showed that a $m$-sheet embedded smoothly and
isometrically into a space of dimension less than $2m$ must have
straight lines in the sheet material which extend across the sheet and
which remain undeformed.  Specifically, there exists through any point
$p$ at least one straight line in the undistorted $m$-sheet $\calS$
which 1) is straight and geodesic in the embedding space $\mathR^d$, and
2) extends to the boundary of $\calS$.  We'll denote this result as
Theorem 1.  
Theorem 1 implies that a $m$-sheet of minimum diameter $L$ cannot be
confined to a $d$ dimensional ball of radius smaller than $L/2$, if $d
< 2m$, where the minimum diameter $L$ is given by
$$
L = 2 \max_{p \in \calS} (\max_{r > 0} \{r : B(p,r) \subseteq \calS\}),
$$
and $B(p,r)$ is the $m$ dimensional ball of radius $r$ centered on $p$.
By taking points far
from the boundary of $\calS$, we may identify lines roughly of the size
$L/2$ or longer.  It is clearly impossible to confine the sheet to a
region smaller than such a line.

Observations of embedded sheets and generalizing the proof from
\cite{shankar.witten} lead us to
conjecture the following extension to Theorem 1.  We conjecture that
through any point $p$ there is a $(2m-d)$-dimensional
subsheet~\cite{foot1}
\begin{enumerate}
\item $R$ is totally geodesic in the sheet.
\item The image of $R$ under the embedding  is
totally geodesic in $\mathR^d$. This together with item 1 implies that the
sheet is {\it flat}.
\item If the point $p$ is a distance $X$ from the boundary of
$\calS$, then the subsheet $R$ through $p$ contains a $(2m-d)$-dimensional
ball of diameter $X$.  
\end{enumerate}

We'll denote these assertions as Conjecture 1. 
The subsheets $R$ can be readily be identified for a simple cone 
in a two sheet.  
For any point $p$ on the cone, $R$ is the half-line extending from the apex 
through $p$. Figure~\ref{fig:butterfly} 
illustrates an example of a 3-sheet in which the 
subsheets $R$ are two-dimensional.  


We may in principle confine an $m$-sheet isometrically 
within a ball of arbitrarily small size by removing
subsets of $\calS$ so that it has ``interior'' boundaries.  We shall denote
the removed part as the {\it defect set} $\calD$.  By removing
sufficiently many subsets, we can assure that all points of the
resulting sheet $\calS'$ are as close to the (interior) boundary as we
like.  
In order that the remaining region be isometric, further conditions are
needed: Conjecture 1 forces some of the removed regions to have a
dimensionality greater than some limit, as we now show. 

\begin{figure}[tbp!]

{

\raisebox{1.8 in}{(a)}
\begin{minipage}[t]{3.0 in}
\center
\epsfig{file=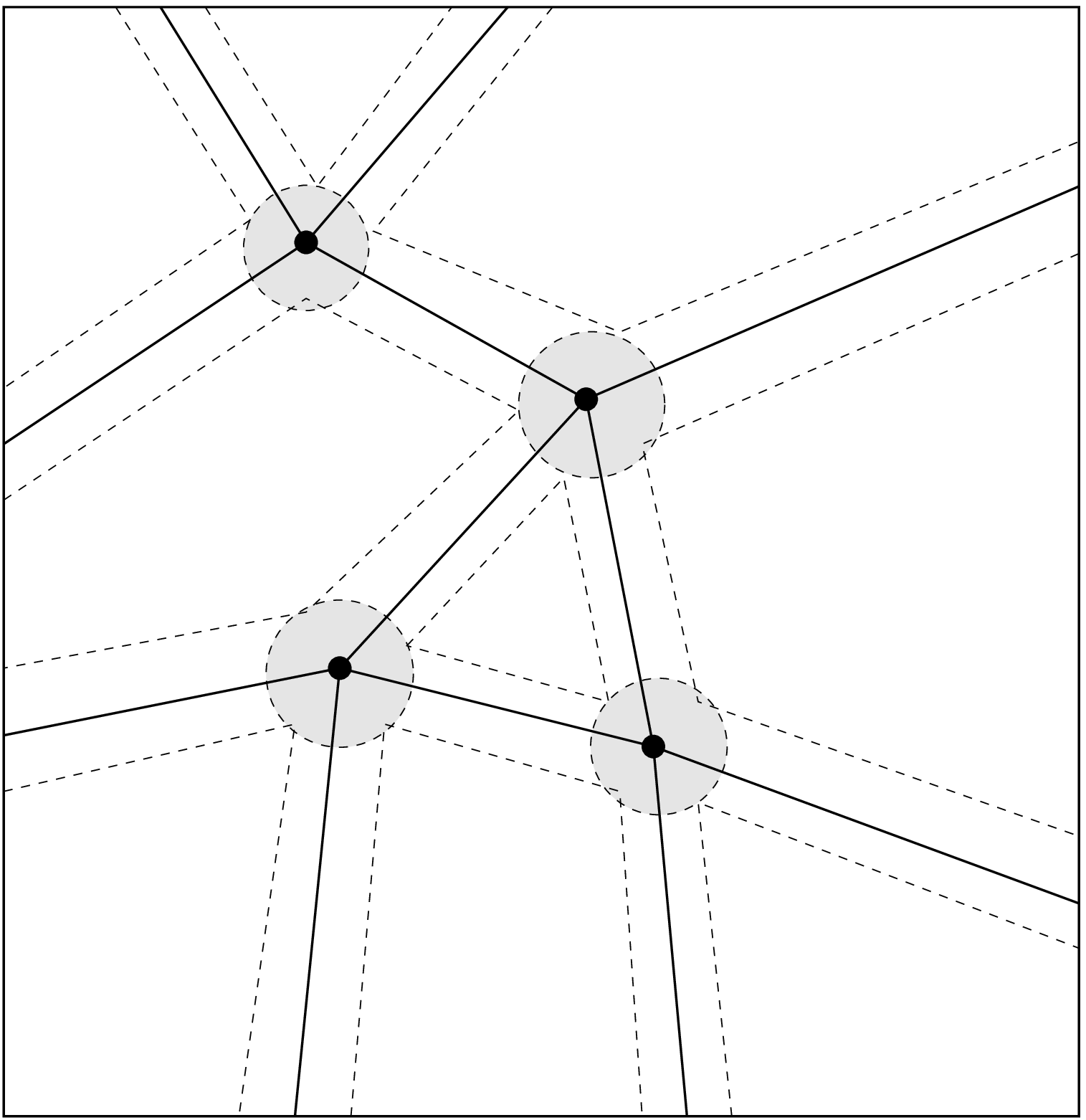, height=2in, width=2in}
\end{minipage}
\hfill
\raisebox{1.8 in}{(b)}
\begin{minipage}[t]{3.0 in}
\center
\epsfig{file=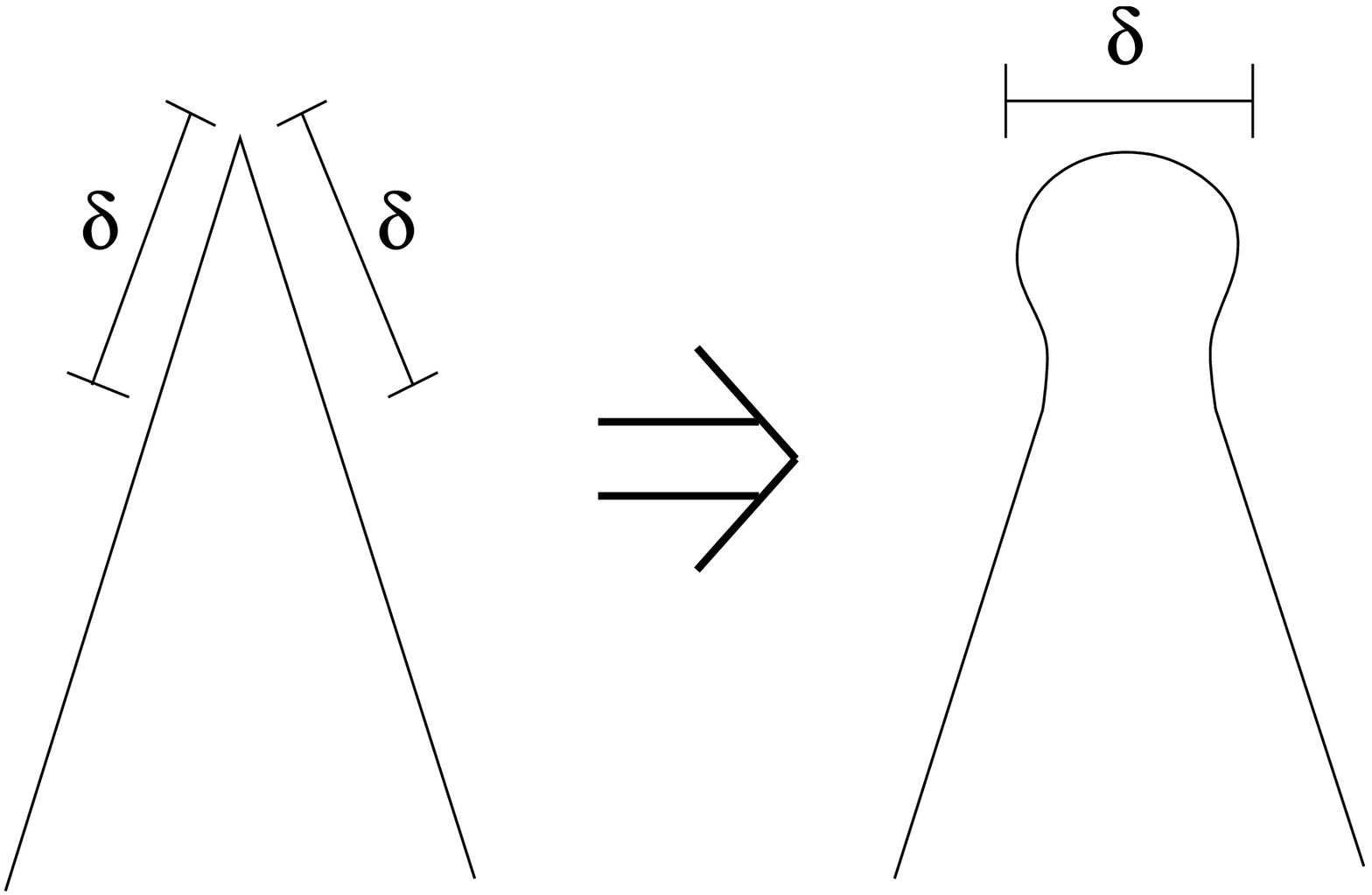, height=1.8in}
\end{minipage}

}

\caption{(a) Illustration of the regions $\calD$, $\calK$, ${\cal
D_\delta}$, and ${\cal K_\delta}$ for a $2$-sheet. The points are a
possible set $\calD$ and the shaded circles are the corresponding 
${\cal D_\delta}$. The solid lines are a possible set ${\cal K - \cal
D}$, and the area within the dashed lines are the corresponding 
${\cal K_\delta - \cal D_\delta}$.
(b) Illustration of a potential way to soften the folding around a
region in $\calK$.}
\label{fig:removal}
\end{figure}

We first confine a convex $m$-sheet $\calS$ within a ball of diameter $X$
much smaller than the minimum diameter $L$ of the sheet.  As indicated
above, this confinement requires strain or singularities.  We now
remove a defect set $\calD$ from the sheet sufficient to allow the
remaining sheet $\calS'$ to be isometric,  as illustrated in 
Figure~\ref{fig:removal}.
We choose a point $p$ further than $X$ from the original $\calS$
boundary, as measured along the sheet.  The subsheet $R$ at point $p$
can have a minimum diameter no greater than $X$; otherwise this flat
subspace would not fit into the confining ball.  Thus the original
boundary of $\calS$ cannot touch the boundary of $R$; $R$ must be bounded
everywhere by $\calD$.  Now, since $R$ is a $(2m-d)$-dimensional set, at
least part of its boundary must have dimension at least $(2m-d-1)$.
(The boundary may also have additional parts of lower dimension, but we
ignore these.)  The set $\calD$ adjacent to this boundary must have at
least this dimension 
as well.  Thus, most $R$'s in the sheet must be bounded over part of
their boundary by defect sets $\calD$ whose dimension is $(2m-d-1)$ or
more.   

These defect sets in strictly isometric sheets have implications for the
confinement of real elastic sheets.  To see this, we repeat the
confinement procedure above taking $\calS$ to be an elastic sheet of
thickness $h$.  We anticipate that regions of concentrated strain will
appear, as they do in ordinary crumpled 2-sheets.  Following the
procedure used above, we remove part of $\calS$ near the regions of
greatest strain, such as the intersection of $R$ and $R'$ in
Figure~\ref{fig:butterfly}. 
Specifically, we remove sets of minimum diameter $\delta$, 
and denote the set of
removed points $\calD_\delta$.  We remove the smallest set
such that the remaining sheet $\calS'_\delta$ becomes 
isometric in the limit as
$h\goesto 0$.  We now reduce the minimun diameter
$\delta$ of our set and repeat
the procedure.  We suppose that the new defect set $\calD_\delta$ is a
subset of the old one, and that we are led to a well-defined limiting
set $\calD$ as $\delta\goesto 0$.  For each $\delta$ we may consider the
boundary of
$R$ for a given point $p$.  Supposing that this boundary also behaves
smoothly, we infer that it retains its dimensionality of at least
$2m-d-1$ inferred above.  Thus the limiting defect set $\calD$ should
also have at least this dimension.  Returning now to the full elastic
sheet $\calS$, we expect the strain to be concentrated on the defect set
$\calD$.  The example of Figure~\ref{fig:butterfly} suggests that $R$
sets are bounded by 
regions of high strain, whose dimension has the minimal value $2m-d-1$.
The numerical work in later sections gives more systematic evidence of
these strained regions.  We shall denote the limiting set $\calD$ as the
{\it strain defect set} and denote each connected part of $\calD$ as a
{\it vertex}.

Although the elastic sheet $\calS'_\delta$ becomes isometric as $h \goesto 0$,
further singularities can develop as $\delta \goesto 0$.  Ordinary 2-sheets in
3-space show this behavior, as illustrated in Figure~\ref{fig:removal}.  
Here the minimal
vertex dimension $2m-d-1$ is 0.  The set $\calD_\delta$ consists of the four
shaded disks: each disk constitutes a vertex.  Removing these disks permits
strain-free confinement to a fraction of the size of the sheet.  However, the
strain-free deformation develops large curvature as $\delta$ becomes small. 
The diverging curvature is concentrated on lines joining the vertices.  Similar
diverging curvature must occur in intact sheets as $h\goesto 0$.  We
denote such regions  by $\calK$, which we call the {\it curvature defect
set}.  
For completeness, we define a set $\calK_\delta$ which contains the regions
of high curvature around $\calK$ for $\delta > 0$.
For intact sheets, we expect the strain to be
significant in the region $\calD$ but very small outside of it -- noting
that the geometric von Karman equation, Eq.~\ref{eq:GvK}, relates 
large gradients in the strain to large curvature, we conclude that  
$\calD$ must be a subset of $\calK$. We
denote each connected 
piece of $\calK - \calD$ as a {\it fold} in the crumpled sheet.
The relationship between these folds and stretching 
ridges~\cite{science.paper} is
discussed in the next section.

	Thus far we have considered effects due to confinement in a
small ball in $\mathR^d$.  We expect similar effects if we impose other
constraints that reduce the spatial extent of the embedded sheet.  We
expect defect sets $\calD$ and $\calK$ 
like those above to form spontaneously \tw 
here as well.  Our numerical investigations reported below do indeed
show such behavior.  We compare our expectations with the numerical
findings in Section~\ref{sec:discussion}.

\begin{figure}[tbp!]
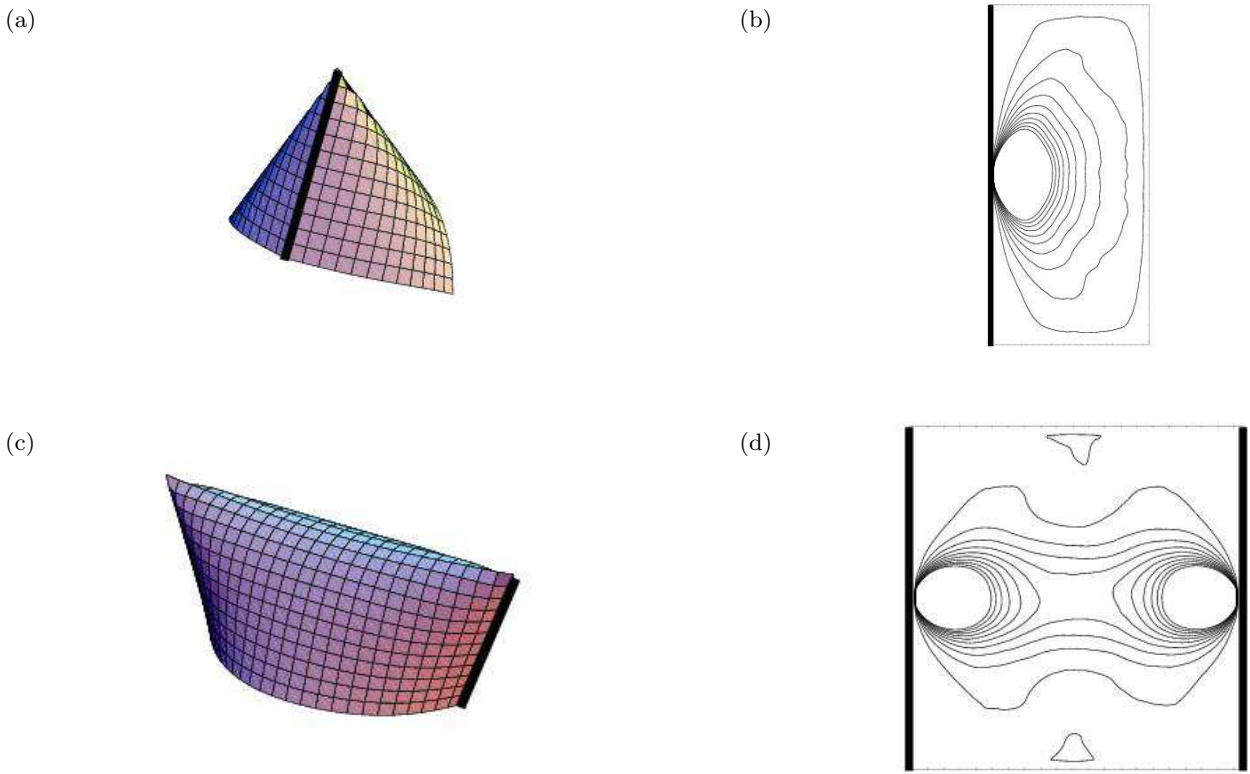


\raisebox{1.8 in}{(a)}
\begin{minipage}[t]{3.0 in}
\center
\epsfig{file=figure3a.eps2, height=2in, width=2in}
\end{minipage}
\hfill
\raisebox{1.8 in}{(b)}
\begin{minipage}[t]{3.0 in}
\center
\epsfig{file=figure3b.eps2, height=2in, width=1in}
\end{minipage}

\vspace{0.2in}
    
\raisebox{1.8 in}{(c)}
\begin{minipage}[t]{3.0 in}
\center
\epsfig{file=figure3c.eps2, height=2in, width=2.5in}
\end{minipage}
\hfill
\raisebox{1.8 in}{(d)}
\begin{minipage}[t]{3.0 in}
\center
\epsfig{file=figure3d.eps2, height=2in, width=2in}
\end{minipage}

\caption{A cone and a ridge formed with disclinations. 
Image (a) 
shows a minimal elastic energy embedding for a $2$-sheet with one 
disclination in $3$-dimensional space. The sheet was $50 \times 100$ 
lattice units in size and had an elastic thickness of $\sim 1/100$ 
in lattice units. 
The disclination was formed by folding one edge at its center point 
and attaching the two halves.
The minimal energy configuration is a cone. Plot (b) 
shows surfaces of equal bending energy for the sheet in (a), plotted
in its material coordinate system. 
Image (c) 
shows a minimal elastic energy embedding for a $2$-sheet with {\it two}
disclinations in $3$-dimensional space. The sheet was $100 \times 100$ 
lattice units in size and had an elastic thickness of $\sim 1/10$ 
in lattice units. The minimal energy configuration is a ridge. Plot (d) 
shows surfaces of equal bending energy for the sheet in (c), plotted
in its material coordinate system. In each image, heavy lines 
indicate edges of the sheet which were joined together to make the
disclinations. 
}
\label{fig:twosheets}
\end{figure}

\section{Energy Scaling}
\label{sec:scaling}

We now return to the consideration of sheets with small
$h > 0$. We consider sheets which are thin enough
that the strains far away from the singular set are much less than
$O(1)$. This is the range of thicknesses that is normally considered in
the study of thin $2$-sheets~\cite{maha,Alex}. In this range
the preferred configuration of a $2$-sheet is well described by 
asymptotically matching 
nearly isometric embedding over most of the sheet to finite boundary
layers around the singular set $\calK$ (within which the strains and
curvatures may become large on the scale of $L$, the manifold size).
We maintain the assumption that energetically 
preferred embeddings will exhibit near isometry outside 
the singular set for $m$-sheets in $d$-dimensional space,
and view the singular set as the subset 
of the material manifold onto which elastic energy condenses as 
$h \rightarrow 0$. 
We wish to study the degree of energy condensation
onto these elastic structures as a function of the material and embedding
dimensions and the elastic thickness $h$. In this section, we show
how the scaling of elastic energy density with volume away from
the condensation regions can be used to quantify the degree
of energy condensation in crumpled $m$-sheets. We distinguish two
cases for the structures involved in crumpling. In the first case, 
$\calK = \calD$ and singular curvature occurs only at
vertices. In the complimentary case, $\calK - \calD \neq
\varnothing$, 
vertices are connected by folds in the $h \rightarrow 0$
limit. We show that these two cases have distictive energy scaling
signatures when $2$-sheets are crumpled in $3$ dimensions. 
Anticipated scaling exponents for general crumpling
are inferred by analogy to lower
dimensional crumpling.

There are three types of data we may use to analyse
minimum energy sheet configurations: the detailed embedding coordinates
and the bending and stretching energy densities in the manifold
coordinates. To see whether energy has condensed in our simulated sheets, 
we first identify regions of high energy concentration
by plotting surfaces of constant bending or stretching elastic
energy in the material coordinates. Fig.~\ref{fig:twosheets} 
illustrates, for the case of a $2$-sheet in $3$ dimensions,
how surfaces of constant bending energy 
highlight the energy-bearing regions of the sheet.
We then look at the coordinate information to associate regions
of energy concentration with either vertices or folds.

For the remainder of our analysis we consider the
energy density data only as a function of volume fraction,
independent of position. This removes any ambiguity in defining the
center points of the high strain regions around $\calD$,
and it provides a natural
framework for defining the degree of energy condensation.
Let the variable $\Phi$ represent the volume fraction in the manifold
coordinates measured from 
the regions of highest to lowest energy density, $0 \le \Phi \le 1$.
Thus, $\Phi$ can be written:
\begineq{phi_def}
\Phi\left(\calL\right) = \frac{1}{V_{total}}
\int_{\calL(\vec{x}') \ge \calL}
d^{m}x',
\end{equation}
where $\calL = \calL_b + \calL_s$ 
is the elastic energy density defined in Eqs~\ref{lame.energy} and
\ref{curvature.energy}, and 
$$
V_{total} = \int_\calS
d^{m}x'.
$$
Surfaces of
constant energy in the manifold coordinates are
also surfaces of constant $\Phi$. 
Inverting Eq.~\ref{phi_def} associates a volume fraction 
with each observed
value of the local elastic energy density.
We can write the 
total energy $E$ in the manifold as
\begineq{totalscaling}
E=\int_0^1 d\Phi' \calL ( \Phi' ).
\end{equation}
We say the energy is {\em condensed} in a volume fraction $\Phi_c$ if for
$\Phi > \Phi_c$ the energy 
density $\calL$ falls away faster than $\Phi^{-1}$. If this is the 
case, then the upper limit of integration can be pushed to infinity 
without changing the value of the integral by more than a 
finite fraction.  By repeating this analysis for the bending energy 
$\calL_b$ or the stretching energy $\calL_s$, we may characterize the 
condensation of these forms of energy individually

We may now make predictions for the elastic energy scaling exponents
based on our knowledge of the 
structures found in crumpled sheets. We first consider the case when the
set $\calK = \calD$. In the familiar crumpling of $2$-sheets in $3$
dimensions $\calK = \calD$ when the sheet contains a single vertex and 
the configuration outside the vertex is conical. In any dimension, 
it is easy to see
that far away from $\calD$ the conformation
should be independent of the small length scale $h$. Far 
away from the curvature singularity
there is no intrinsic length scale, so 
simple dimensional analysis tells us that the curvature must be a 
numerical multiple of $r^{-1}$, where $r$ is the distance from the
vertex. The preferred embedding is thus a simple generalization of a
cone, with straight line generators radiating from a central vertex and
transverse curvatures decreasing as $1/r$ along the generators.
The cone configuration is isometric outside the vertex for $h=0$,
but for $h > 0$ it  acquires small but finite strain.
Dimensional analysis of the force von Karman equation, 
Eq.~\ref{eq:FvK}, for a curvature of the form $C(r, \theta) =
g(\theta)/r$ yields strain scaling of the form $h^{2}/r^2$ for
nearly isometric embeddings. 
Thus, for energetically preferred embeddings,
the bending and stretching energy
densities should scale as $1/r^2$ and $h^{2}/r^4$ respectively.
We can express this energy scaling in terms of the volume fraction
$\Phi$ by finding how $\Phi$ grows with $r$. 
In any principal material direction, smooth curvature of order 
$C$ will typically persist over a length of order $1/C$. 
In a $3$-sheet, the volume of high
energy density surrounding an energetic cone generator
will therefore grow as $r^2$ if the curvature in one 
material directions transverse to the generator dominates, 
or as $r^3$ if the curvatures in both transverse directions are on the
same order.
The bending energy density will respectively
scale as $\Phi^{-1}$ or $\Phi^{-2/3}$. 
Since the strain along a generator dies twice as quickly as the curvature, 
the stretching energy density will correspondingly scale 
as $\Phi^{-2}$ or $\Phi^{-4/3}$.
Thus we surmise that conical scaling has the typical form
\begineq{cone_scaling}
\begin{array}{c} 
\calL_b \sim \Phi^{-p} \\ 
\calL_s \sim \Phi^{-2p}, \\
\end{array} 
\end{equation}
where $p=2/(1+n)$, and $n$ is an integer equal to the number of 
transverse curvature directions along energetic cone generators.
In all the geometries accessible to $2$ or $3$-sheets, 
the stretching energy is condensed while the bending is
not -- this is not the case in all higher 
dimensions, where the value of $n$ can 
be greater than $3$ and the stretching energy not condensed.

We now consider configurations that 
have $\calK - \calD \neq \varnothing$.  We have denoted each
connected piece of $\calK - \calD$ as a {\it fold} in the previous 
section.  At this point we need to make a distinction 
between folds and ridges.
For $2$-sheets in $3$ dimensions with
$h>0$, folds have an energetically preferred local structure.
We describe folds in this context as ridges, a term which 
encompasses both the geometric and energetic structure.
In general crumpling, we don't know {\em a priori} whether the local 
structure around folds will be similar to that in lower 
dimensional crumpling, so we must make our definition of 
a ridge more precise. Since we already have a geometrical 
descriptor for folds, we use the term ridge to describe a 
certain kind of energetic structure associated with folds.
The generalization of a ridge is a boundary layer around a fold
whose energy scaling depends on two length scales --- the elastic 
thickness $h$ of the sheet and the length $L$ of the fold. 
Clearly in the thin limit these are the only two length scales which 
can be important around the fold. Conversely, there must be 
at least two length scales if there is to be any non-trivial scaling 
of the ridge profile
with thickness $h$. The presence of two length scales allows for a 
balance between the coupled bending and stretching energies, which is 
also a hallmark of ridges (and could be used as an alternative 
equivalent definition)

For $2$-sheets in $3$ dimensions, the condition $\calK - \calD \neq 
\varnothing$ occurs \eg when 
there are two vertices joined by a ridge, as in 
Figure~\ref{fig:twosheets}.
It is well known~\cite{Alex} that the elastic energy density in the 
region surrounding $\calK$ which encompasses a ridge is less than that
at vertices (the region surrounding $\calD$) 
but much greater than that in the region of $\calS-\calK$ 
away from the energy condensation structures. Ridges begin and 
end at vertices, with the elastic energy density falling smoothly along the 
ridge length away from each vertex.
This implies that when ridges are present, the
scaling of $\calL(\Phi)$ at values of $\Phi$ much less than $1$
but large enough to 
fully contain the vertices will 
be determined by the parts of ridges which are closest to vertices.
Ridges are also known to
have a complicated spatial structure,
but we assume that the ridge solution converges to a 
simple scaling solution near the vertex, where the ridge length should
become unimportant. 
It has been shown~\cite{science.paper,Alex} that the total bending and
stretching energy of ridges in $2$-sheets scale the same way with 
manifold length scales and obey a virial theorem: the ratio of 
the total bending to stretching energy is 1 to 5.
The same virial ratio was also demonstrated for ($m-1$)-dimensional ridges 
in $m$-sheets~\cite{eric}.
We therefore infer that to lowest order, the bending and stretching 
energy densities must follow {\em identical} scaling in the simple scaling 
region near vertices.
The virial relation, $E_{b}=5 \times E_{s}$, should
also be evident in the ratio of scaling prefactors
for the two energies. Furthermore, the total elastic energy in a
ridge diverges as the length of the ridge becomes 
infinite~\cite{science.paper}, so the
energy density along the ridge should not fall faster than $\Phi^{-1}$. 
Thus we expect the
scaling behavior of ridges to follow
\begineq{ridge_scaling}
\left. \begin{array}{c} \calL_b \sim \Phi^{-q} \\ 
\calL_s \approx 1/5 \calL_b \sim \Phi^{-q} \\
\end{array} \right\} 0<q<1.
\end{equation}

This dependence implies that strain energy has not condensed onto the
vertices alone if ridges are present. In general, our assumption of
near-isometry away from the defect set implies
that strain will condense out of the bulk of the $m$-sheet. Thus we
expect that the strain must condense onto the ridges and vertices --
onto the full set $\calK$.
This means that, beginning at some 
$\Phi_c < 1$ which marks the
boundary of the ridge scaling region, 
there will be a more rapid drop-off (faster than $\Phi^{-1}$)
of strain energy with volume away from the ridges.

We can calculate the anticipated scaling exponent $q$ above based on the 
anticipated scaling of the ridge width $w(r)$ at a distance $r$ from a 
vertex, \viz $w(r) = w(X) f(r/X)$, where $X$ is the length of the 
ridge.  Previous work \cite{eric.math} shows that $w(X) \sim h
(X/h)^{2/3}$ for $m-1$-dimensional ridges in $m$-sheets.  
We anticipate that $w(r) \ll w(X)$ when $r \ll X$, and 
that in this regime $w(r)$ is independent of $X$.  Then our scaling 
assumption implies $w(r) \approx h (r/h)^{2/3}$.  The transverse 
curvature $C(r)$ is as usual presumed to be of order $1/w(r)$.  Lobkovsky 
\cite{Alex} originally derived this scaling property based on more detailed 
assumptions. The curvature
energy should be significant in a region of width $w$ around the center
of the ridge. The local energy
density therefore scales as $C^2 \sim r^{-4/3}$, while the high
energy volume should grow as $\Phi \sim r \times 1/C =
r^{5/3}$. Thus the above curvature scaling leads
to $\Phi^{-4/5}$ scaling for both $ \calL_s $  and $ \calL_b $
around the vertex if it is the end-point of a ridge. This scaling was
originally
derived for $2$-sheets embedded in $3$ dimensions. Other
work~\cite{eric.math} suggests that the same scaling should hold for
$m$-sheets in ($m+1$)-dimensional spaces. 
For $m$-sheets, ridges with spatial extent $X$ in $l$ long 
directions and width of the form $w(x)$ given above in the 
remaining $m-l$ directions will occupy a total volume
$X^m (h/X)^{(m-l)/3}$. Compared with the total volume of the 
manifold, which is of order $X^{m}$, the  high energy volume fraction is  
$\Phi_c \sim (h/X)^{(m-l)/3} \ll 1$.
Thus there is energy condensation onto ridges.
For general dimensions, we 
reason that any balance of bending and stretching energies should lead 
to a virial relation, and a virial relation in turn implies parallel 
scaling of the two energy densities. So, Eq.~\ref{ridge_scaling} 
should hold for all higher dimensional generalizations of ridges.


\section{Numerical Methods}
\label{sec:numerics}

For the present study we have generalized the numerical 
approach of Seung and Nelson\cite{Seung.Nelson}, modelling
an $m$-sheet as an $m$-dimensional rectangular lattice and
adding terms to the elastic energy to produce bending stiffness. 
Properly speaking, we simulate phantom $m$-sheets,
which can pass through themselves. In the latter part
of this section we discuss the parameter range in which the 
phantom $m$-sheet behaves like a physical $m$-sheet, 
as well as the special
implications of the phantom approximation on the structure of vertex
singularities.

Our manifold is a hypercubic array of nodes labeled by
$\ulI \definedas \{i_1, ...,
i_m\}$.  Each node has a $d$-dimensional position vector $\vec{r}
(\ulI)$.  The relaxed lattice has a nearest-neighbor distance
$a$.  The lattice displacement from a site at $\ulI$ to a nearby one can be
expressed by a vector of $m$ integers, 
$\ulDelta$.  It is convenient to define
the lattice displacements $\vec{u}(\ulI,\ulDelta)$,  
defined as the displacement between the node at site
$\ulI$ and the one shifted by $\ulDelta$ \begineq{define.u} \vec{u}
(\ulI, \ulDelta) \definedas -\vec{r}(\ulI) + \vec{r}(\ulI+
\ulDelta) .\end{equation} The stretching energy $U (\{\vec{R}\})$ for a
$3$-sheet ($m=3$) is now defined as
\begin{eqnarray}
\label{lattice.stretch.energy}
U & \definedas & G \sum_{\ulI} 
\sum_{\ulDelta=nn}(|u(\ulI, \ulDelta)| -
a)^2   \nonumber \\
& &  + c_s  \sum_{\ulI} \sum_{\ulDelta=nnn}(|u(\ulI, \ulDelta)| -
\sqrt{2} a)^2 .
\end{eqnarray} 
Here the $nn$ denotes the six
nearest-neighbor sites $\ulDelta=(\plusorminus 1, 0, 0), (0, \plusorminus 1, 0)$,
and $(0, 0, \plusorminus 1)$.  The $nnn$ sites are the 12 second
neighbor sites of the form $(\plusorminus 1, \plusorminus 1, 0)$, etc.
The weight coefficient $c_s$
assures that $U$ is isotropic: \ie independent of
the direction of strain relative to the lattice.  We found by direct 
calculation of the elastic energy for uniform strain in the $(1,0,0)$,
$(1,1,0)$ and $(1,1,1)$ directions, minimized with respect to lateral
expansion, that $U[\gamma]$ was equal for the three directions of strain
when $c_s = 1$. The corresponding Poisson ratio
is $1/4$.
By expanding Eq. (\ref{lattice.stretch.energy}) for small deviations of
the $3$-sheet from zero deformation
and equating terms with those of
Eq.(\ref{lame.energy}),
we infer that for our lattice $\mu= 4 a^2 G$ and $\lambda = \mu$.  

We use a discrete form of Eq.~\ref{curvature.as.derivative}
to determine the curvatures in our
simulated $m$-sheet.  
For each origin site $\ulI$ we evaluate 
the diagonal elements $\vec{\kappa}_{ii}$ from the nearest-
neighbor separations:
\begineq{diagonal.curv}
\vec{\kappa}_{ii} \approx \frac{1}{a^2} \left[ \left(
\vec{r}(\ulI + \ulDelta_i)-\vec{r}(\ulI) \right)-
\left(\vec{r}(\ulI)-\vec{r}(\ulI - \ulDelta_i) \right) \right]
\end{equation}
The off-diagonal elements 
$\vec{\kappa}_{ij}$, $i \ne j$ are computed in similar fashion
from the next nearest-neighbor positions:
\begin{eqnarray}
\label{off.diagonal.curv}
\vec{\kappa}_{ij} & \approx & \frac{1}{4 a^2} \left[ 
\left(
\vec{r}(\ulI + \ulDelta_i + \ulDelta_j)-
\vec{r}(\ulI + \ulDelta_i - \ulDelta_j) \right) \right. \nonumber \\
& & - \left. \left(
\vec{r}(\ulI - \ulDelta_i + \ulDelta_j)-
\vec{r}(\ulI - \ulDelta_i - \ulDelta_j) \right)
\right]
\end{eqnarray}

Once the curvature matrix $\vec{\kappa}$ is known for each site, we may
compute the curvature energy $E_b$ from Eq.(\ref{curvature.energy}).
To save computational time we do not project the
curvature vectors onto the normal space of the manifold
at $\ulI$. This amounts to including tangential components in
the curvature tensor defined in
Equation~\ref{curvature.as.derivative}. It is the usual practice in
linear elasticity to neglect these terms because of their
smallness~\cite{Lame.ref}, so leaving them in for computational
efficiency does not introduce any significant change to the energy
density profile.


The sizes and elastic thicknesses of the lattices used in our
simulations were arrived at through a trial-and-error balancing
of computational resources and data quality. We minimized elastic
energy using an inverse gradient routine, which theoretically 
converges in $\sim N^2$ steps for a harmonic potential with $N$
degrees of
freedom~\cite{conj_grad}. 
However, experience shows that the convergence becomes much
less efficient when we make the elastic sheets very thin, since 
in this limit the 
total energy functional is highly non-linear and has large prefactors
for the highest-order terms. This effect in elastic simulations was
described in~\cite{Cerda}, but their ``reconditioning'' approach to 
regaining fast convergence requires too much computational overhead
to be of use on large $3$-dimensional lattices. The computational cost
of larger lattices must be balanced against the range of validity of the
discrete lattice approximation. The lattice can only accurately
accommodate embeddings where the radius of curvature, $1/C$, is locally 
much greater than the spacing between lattice points. We have no hope of
maintaining accuracy at a vertex, which is a near singularity, but we
try to stay within an operating range where the sharpest features away
from vertices have radii of curvature at least a few times the 
inter-lattice point spacing. This indirectly constrains the 
thickness of the
elastic manifold we simulate, since features become sharper 
as the manifold is made thinner.

Our standard simulational procedure was to begin with a lattice about
$30$ units on a side, since this was the smallest lattice where fine
features were clearly visible. After the elastic energy of the manifold
was minimized on this lattice, we interpolated the result onto an
80 unit lattice and minimized again. Then we decreased the elastic
thickness of the manifold on the larger lattice
over a process of several minimizations. When
the elastic thickness of the manifold becomes very small, the material
becomes prone to falling into broad local minima with
fine-scale crumpling which confuses the energy data. Slowly decreasing the
thickness is a method to avoid this fine-scale crumpling. In most of
the following sections we
present the result of simulations on $80$-unit lattices with an
elastic thickness of $\approx 0.02$ lattice units. The entire process of
generating each minimized lattice took up to several weeks on a 233 MHz,
Pentium II based linux computer using a gcc compiler.

We note that our lattice simulates
a phantom $m$-sheet, which can pass
through itself without penalty. Since the energetic properties we study
follow from local laws, and we stay in a thickness regime where
curvature is non-singular almost everywhere, the fact that our sheets
are not self-avoiding does not affect the conclusions we draw from our
data. The effect of the phantom $m$-sheet behavior on the dimensionality
of vertex structures, where curvature does become singular, is discussed
in Section~\ref{sec:discussion}. 
In particular, 
as the thickness goes to zero, the minimum energy configurations need
not converge to objects that have the local structure of a manifold.
For example, in the
vicinity of a vertex in a phantom sheet, as $h \rightarrow 0$, the
configuration might converge to a branched manifold 
({\it e.g.}, a cone which
winds twice around some axis).

For geometries which generated several disclinations in a single
manifold, we took special care with initial conditions to insure that the
system moved towards a symmetric final state. Relaxed states which
contained a collection of identical vertices gave much cleaner scaling
and always had a lower total energy than those which contained an
ensemble of vertices with different local structure.
Thus, when we started the sheet in a state with several folds, we
separated the opposite sides of the folds slightly in globally symmetric
ways to determine how they would relax.


\begin{figure}[tbp!]
\center
\epsfig{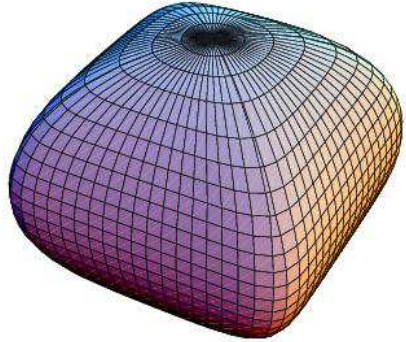}

\caption{Equipotential surface of the spatially
confining potential for a square
$2$-sheet 
embedded in 3 dimension. }
\label{fig:cubicpot}
\end{figure}

\begin{figure}[tbp!]
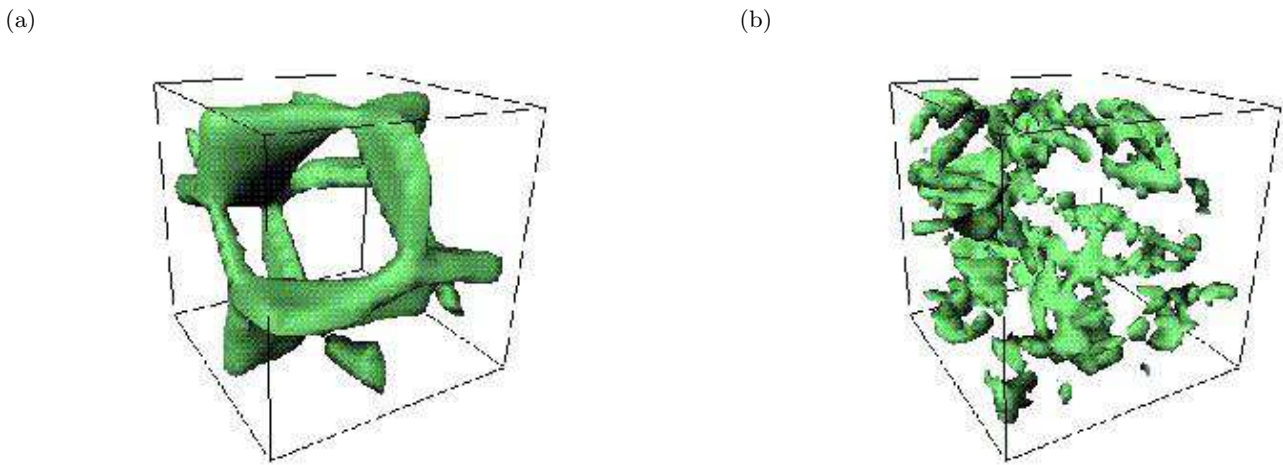

\center
\small
\raisebox{2.3 in}{(a)}
\begin{minipage}[t]{3.0 in}
\centering
\epsfig{file=figure5a.eps2, height=2.5 in, width = 2.5 in}
\end{minipage}
\hfill
\raisebox{2.3 in}{(b)}
\begin{minipage}[t]{3.0 in}
\centering
\epsfig{file=figure5b.eps2, height=2.5 in, width = 2.5 in}
\end{minipage}

\caption{ Energy condensation map for spatially
confined cubes.  The cubes
were $X=20$ lattice sites wide and had elastic 
thickness $h = .075 X$. 
Image (a) shows a surface of constant bending
energy density in the material coordinate system for a cube
embedded in $4$ dimensions. The surface encloses the $\approx 10\%$ volume
fraction with the highest energy concentration. 
Image (b) shows a surface of constant bending
energy density for a cube embedded in $5$ dimensions.
This surface encloses the $\approx 7\%$ volume
fraction. 
The wireframes represent the edges of the cubes' material 
coordinates.}
\label{fig:conf3in45_ede}
\end{figure}

\begin{figure}[tbp!]
\center
\small
\raisebox{1.3 in}{(a)}
\begin{minipage}[t]{1.5 in}
\centering
\epsfig{file=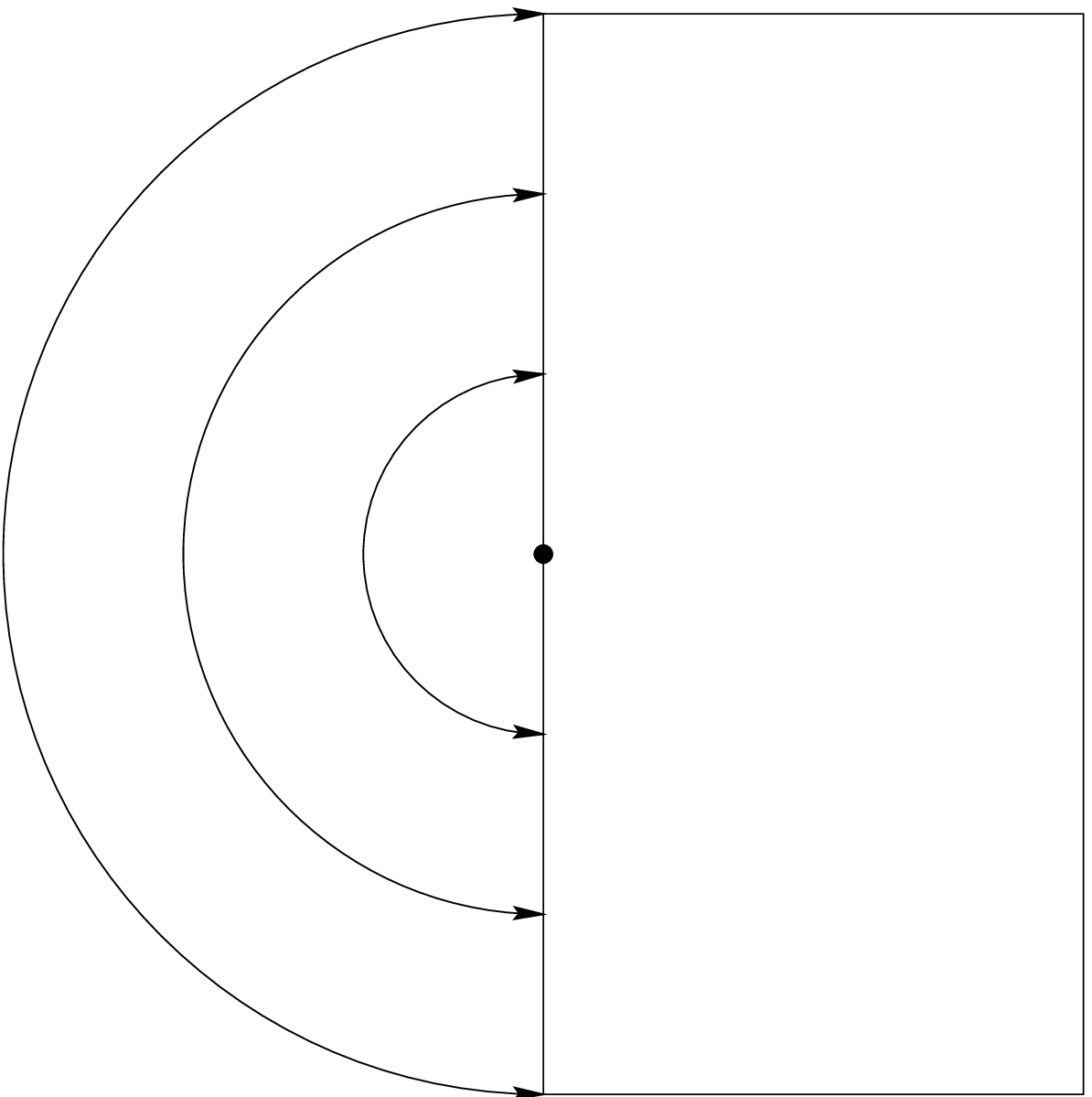, height=1.2 in, width = 1.2 in}
\end{minipage}
\hfill
\raisebox{1.3 in}{(b)}
\begin{minipage}[t]{1.5 in}
\centering
\epsfig{file=figure6b.eps2, height=1.5 in, width = 1.5 in}
\end{minipage}
\hfill
\raisebox{1.3 in}{(c)}
\begin{minipage}[t]{1.5 in}
\centering
\epsfig{file=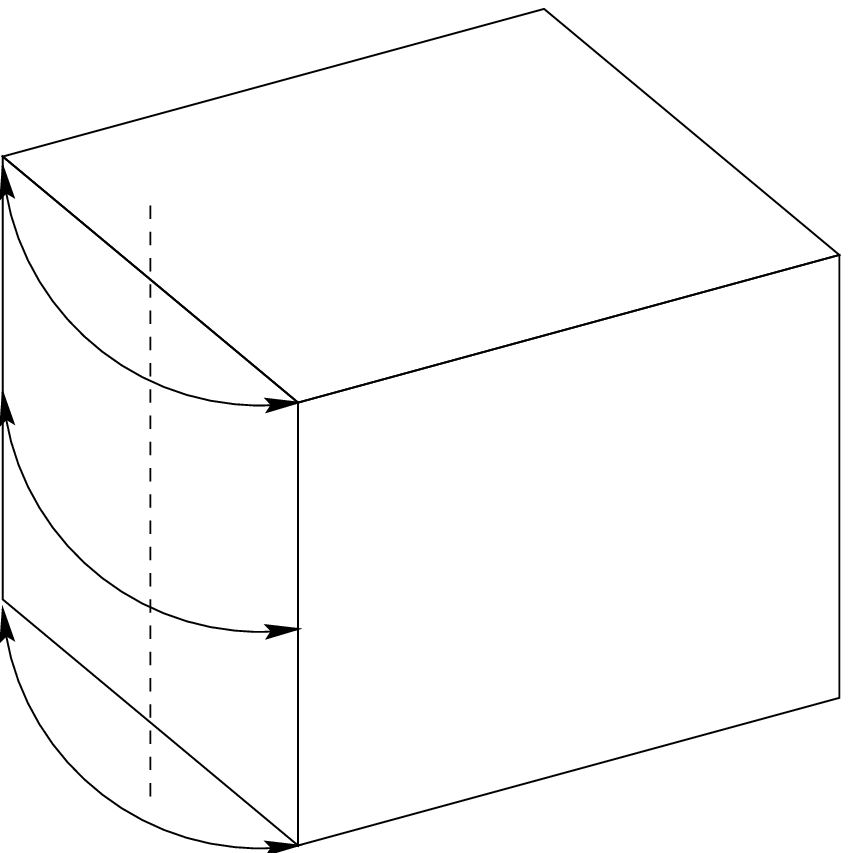, height=1.5 in, width = 1.5 in}
\end{minipage}

\caption{Method for creating disclinations. In (a), one edge of
$2$-sheet is folded and attached as shown. Points on the edge are
identified, but the curvature is not continued across the 
seam. Image (b) shows 
an equilibrium configuration of a $2$-sheet
constructed as in (a) after elastic energy minimization
(as in Fig~\ref{fig:twosheets}). Illustration
(c) shows how the same technique is used to make a line disclination in
an cubic $3$-sheet.
}
\label{fig:makedisclin}
\end{figure}

\section{Spatially Confined Sheets}
\label{sec:confined}

In this and the next three sections we report the results of our
simulations. In this section we explore 
the distortions resulting from spatially 
confining our
3-sheet in a contracting volume.
Spatial confinement was simulated
with an $(R/b)^{10}$ potential, which acts essentially
like a hard wall at a radius $b$. 
We tailored the potential to our cube-shaped $3$-sheet by making
the equipotential lines nearly cubical in three
spatial dimensions and spherically symmetric in all remaining 
dimensions. This reduced edge effects at the corners of our
cubes.
The exact form of the potential was
\begin{equation}
E \propto \sum_{i=1}^{3} \left[ \left(\frac{x_i}{b}\right)^2 + 
\sum_{j=4}^d \left(\frac{x_j}{b}\right)^2 \right]^5.
\end{equation}
An equipotential surface of this potential 
for a $2$-sheet in $d=3$ is shown in
Fig.~\ref{fig:cubicpot}.

We began our simulations with the hard wall potential just outside
the boundaries of the cube, then progressively moved the walls inward
on all sides until the geometrically 
confining volume had only half the spatial
extent of the resting cube in any direction. 
The value of $b$ was decreased in ten
equal steps, with the lattice
allowed to relax to an elastic energy minimum
after each step. This procedure simulates a gentle
confinement process, which allows applied stress to propagate 
through the entire manifold volume instead of being caught in
a strong ridge network at the outside edges. Gentle confinement is
essential to good convergence of the inverse gradient routines
used to minimize the elastic energy.

Since the spatial 
confinement technique requires multiple inverse gradient
minimizations for each simulation, it is not computationally
practical to run on large grids. Also, the data obtained from this 
method does not lend itself as well to numerical analysis, since
the energy gradient from the tails of the hard wall potential
mix with the elastic energy densities. Still, simulations performed on 
smaller grids (20 lattice units extent) show
some interesting qualitative differences between confinement in
$d=4$ versus $d=5$.
As Fig.~\ref{fig:conf3in45_ede} shows, the regions of highest energy
density are well organized line-networks for $d=4$ but are much more
scattered and disorganized for $d=5$. 
Our arguments for the minimal dimensionality of $\calD$
presented in Section~\ref{sec:isometric}  predict a minimal
dimension of $1$ and $0$ respectively for $3$-sheets embedded in 
$4$ and $5$ dimensions. 
If we assume that the high energy
regions seen in Fig.~\ref{fig:conf3in45_ede}
surround parts of $\calD$, then the qualitative
data supports these values for $\dim \calD$.


\section{Disclination Pairs}
\label{sec:disclin}

To gain a better understanding of elastic energy condensation in
$m$-sheets,  we analyze
several simpler forms of distortion. We first study pairs of
disclinations. One way to create a disclination in a square
$2$-sheet is to join two adjacent corners and the edge connecting them,
as shown in Fig.~\ref{fig:makedisclin}(a).
The disclination relaxes into a conical shape like that shown
in Fig.~\ref{fig:makedisclin}(b).
Placing two or more conical disclinations in a $2$-sheet
in $d=3$ causes ridges to form which are apparently equivalent 
to those connecting vertex singularities in a confined sheet.
A corresponding technique for creating folds in a $3$-sheet is
to add line-like wedge disclinations into the manifold.
We simulate line disclinations in $3$-cubes numerically by folding
faces of an elastic cube down the center and connecting the
two halves as shown in Fig.~\ref{fig:makedisclin}(c).

It can be shown by construction that
the $3$-cube in embedding space 
$d>3$ can accommodate one line 
disclination without stretching. One can construct such
an embedding by bending each plane perpendicular to
the line disclination into an identical cone. 
However, pure bending configurations for a $3$-cube with
two such line-disclinations will in general require folds.
It is energetically favorable for the
cube to stretch to avoid singular curvatures,
so we may expect 
the sheet to form ridges with
the same degree of elastic energy condensation as in a 
physically confined $3$-sheet.

We begin our study of disclinations in general dimensions by 
simulating a $2$-sheet with two sharp bends embedded in either 
$3$ or $4$-dimensional space. From previous work~\cite{science.paper}
the $2$-sheet in $3$ dimensions should
from a simple ridge -- its expected 
behavior in $4$ dimensions is not known.
Next we turn our attention to $3$-sheets, beginning
with a  simulation of a half-cube with a single line disclination embedded
in either $4$ or $5$ dimensions.
This simulation will
verify the predicted scaling of a simple cone.
Then, to induce elastic energy condensation 
we construct $3$-cubes with 
two line disclinations at opposite cube faces
and embed them in $4$ or $5$ dimensions.
Since there is no guarantee
that our procedure will find the global energy minimum, we start the
cubes in many different initial conditions.
We investigate the behaviors when the line 
disclinations are either parallel or perpendicular to one 
another in the material coordinates.

\begin{figure}[tbp!]

\center
\small
\raisebox{1.9 in}{(a)}
\begin{minipage}[t]{3.0 in}
\centering
\epsfig{file=figure7a.eps2, width = 3.0 in}
\end{minipage}
\hfill
\raisebox{1.9 in}{(b)}
\begin{minipage}[t]{3.0 in}
\centering
\epsfig{file=figure7b.eps2, width = 3.0 in}
\end{minipage}

\raisebox{1.9 in}{(c)}
\begin{minipage}[t]{3.0 in}
\centering
\epsfig{file=figure7c.eps2, width = 3.0 in}
\end{minipage}
\hfill
\raisebox{1.9 in}{(d)}
\begin{minipage}[t]{3.0 in}
\centering
\epsfig{file=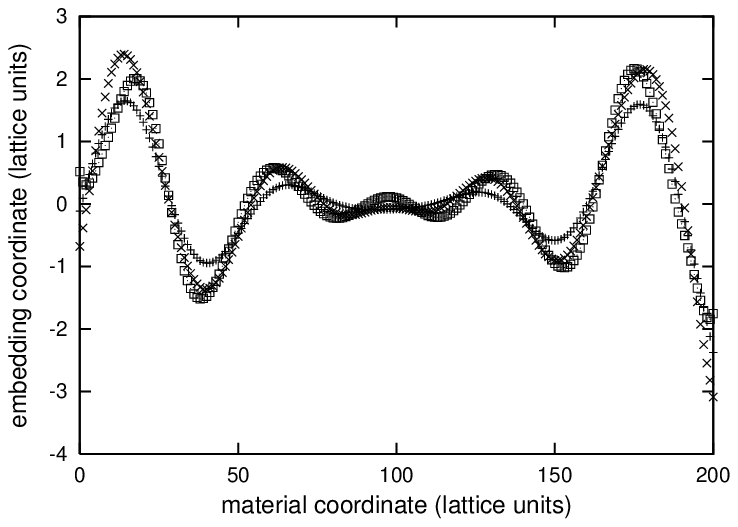, width = 3.0 in}
\end{minipage}

\caption{Equilibrium embedding coordinates for elastic $2$-sheets with
two $90^o$ bends. The sheets
were $X=200$ lattice sites wide and had elastic 
thickness $h = 2 \times 10^{-4} X$. The bends were imposed by
attaching opposite edges to a rigid right angle frame. Image (a) shows
the three embedding coordinates for a $2$-sheet in $3$-dimensional
space. Image (b) shows the same three embedding coordinates for a 
$2$-sheet in $4$-dimensional space. Image (c) shows the fourth embedding
coordinate (not shown in (b)) 
for the $2$-sheet in $4$-dimensions, plotted against the
sheet's material coordinates. In (c), the value of the embedding
coordinate has been multiplied by $20$ to enhance contrast.
In (d) the embedding coordinate shown in (c) is plotted against material
coordinate down the folding line for three different material
thicknesses. The (+) symbols correspond to an elastic thickness of
$1$ lattice unit, the 
($\times$) symbols correspond to an elastic thickness of
$0.1$ lattice unit, and 
the ($\Box$) symbols correspond to an elastic thickness of
$0.01$ lattice unit.
}
\label{fig:2sheet-embed}
\end{figure}

\begin{figure}[tbp!]
\center
\small
\raisebox{1.9 in}{(a)}
\begin{minipage}[t]{3.0 in}
\centering
\epsfig{file=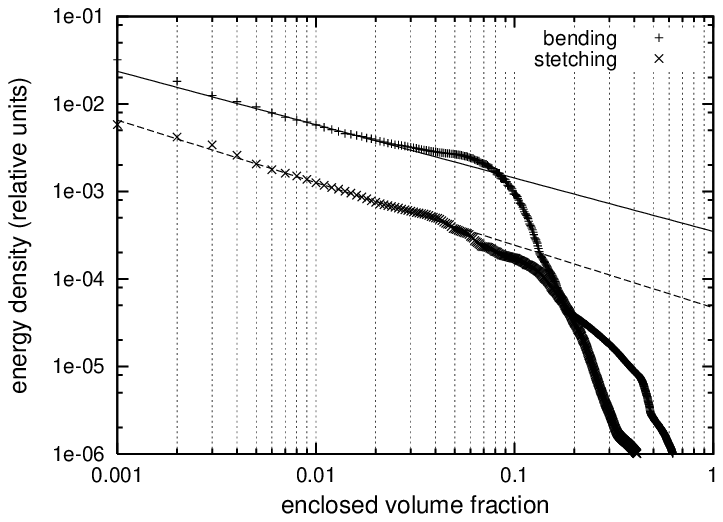, height=2.1 in, width = 3.0 in}
\end{minipage}
\hfill
\raisebox{1.9 in}{(b)}
\begin{minipage}[t]{3.0 in}
\centering
\epsfig{file=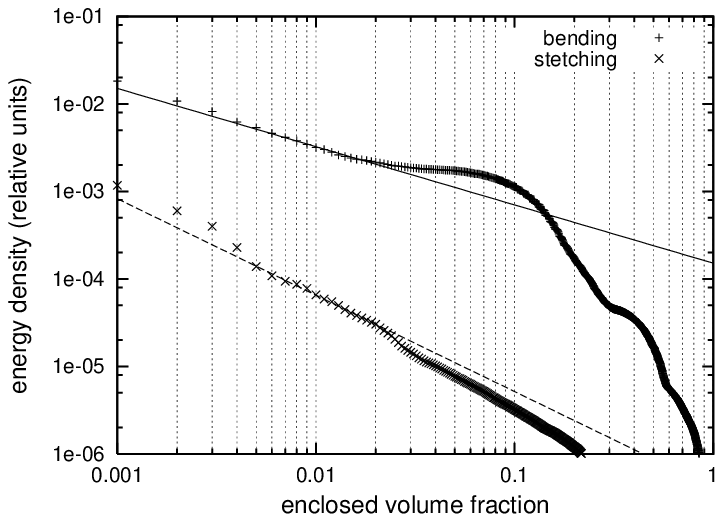, height=2.1 in, width = 3.0 in}
\end{minipage}

\caption{Energy density plots for elastic $2$-sheets with
two sharp bends. The sheets
were $X=200$ lattice sites wide and had elastic 
thickness $h = 2 \times 10^{-4} X$.  
In each graph the $+$ symbols denote 
bending energy while the $\times$ symbols denote stretching energy.  
Energies are expressed in arbitrary units.  
Horizontal axes are area fraction $\Phi$.  Graph (a) shows 
local stretching
energy density $\calL_s$ and bending energy density $\calL_b$
versus area fraction $\Phi$ at or above this energy density
from an embedding in 3 dimensions.  
Graph (b) shows the same quantities from 
an embedding in 4 dimensions.  
In both graphs the straight lines are power law fits to the 
bending and stretching energy densities.
In all graphs the energy fits are to the region 
between $0.5 \%$ and $2.0 \%$ volume fraction.
In (a), the solid line is a fit to the 
bending energy density, with scaling exponent $-0.61$, and
the dashed line is a fit to the 
stretching energy density, with scaling exponent $-0.71$.
In (b), the solid line is a fit to the 
bending energy density, with scaling exponent $-0.66$, and
the dashed line is a fit to the 
stretching energy density, with scaling exponent $-1.10$.}
\label{fig:2sheet-hist}
\end{figure}

\begin{figure}[tbp!]
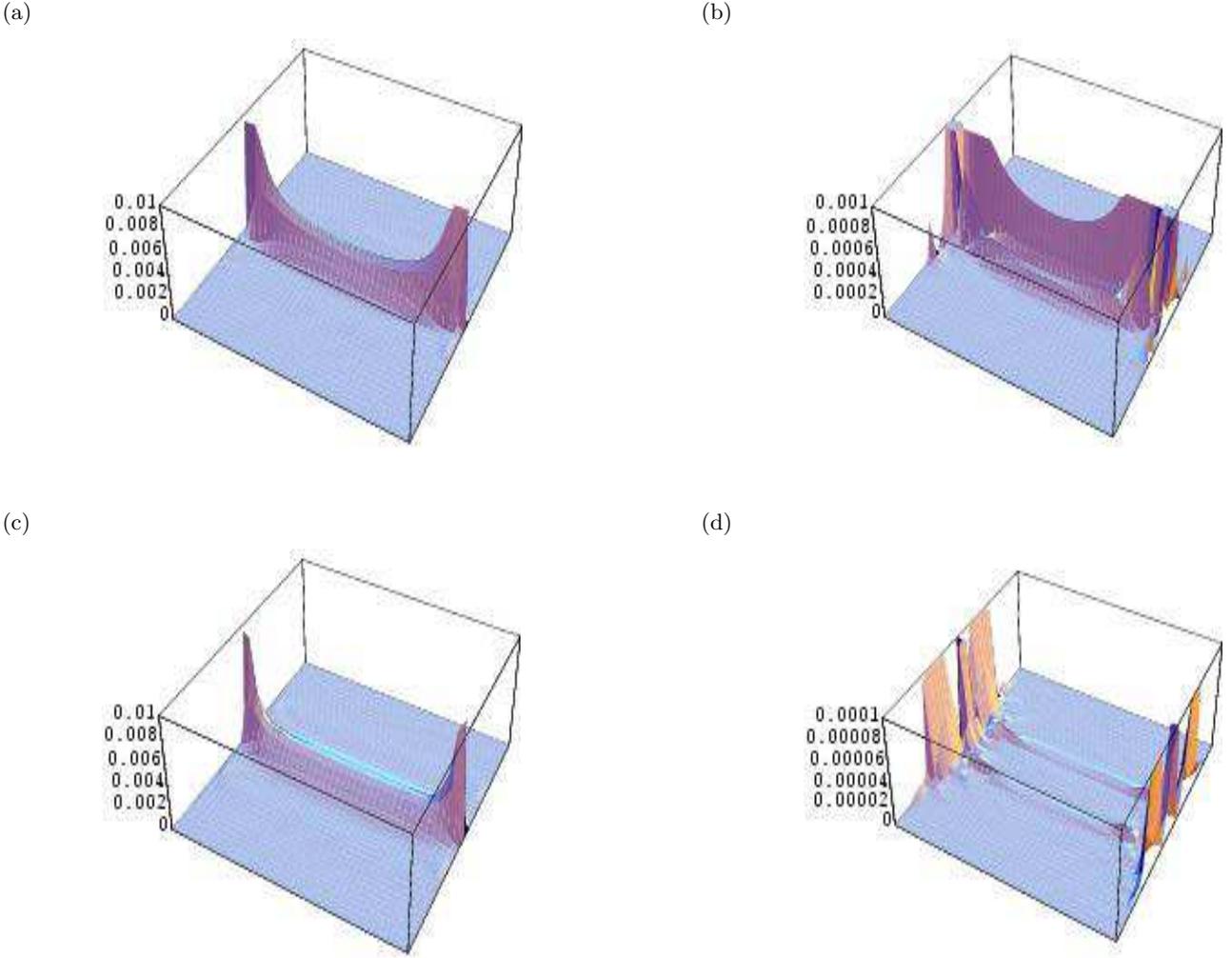

\center
\small
\raisebox{2.5 in}{(a)}
\begin{minipage}[t]{3.0 in}
\centering
\epsfig{file=figure9a.eps2, height=2.5 in, width = 2.3 in}
\end{minipage}
\hfill
\raisebox{2.5 in}{(b)}
\begin{minipage}[t]{3.0 in}
\centering
\epsfig{file=figure9b.eps2, height=2.5 in, width = 2.3 in}
\end{minipage}

\vspace{0.2in}

\raisebox{2.5 in}{(c)}
\begin{minipage}[t]{3.0 in}
\centering
\epsfig{file=figure9c.eps2, height=2.5 in, width = 2.3 in}
\end{minipage}
\hfill
\raisebox{2.5 in}{(d)}
\begin{minipage}[t]{3.0 in}
\centering
\epsfig{file=figure9d.eps2, height=2.5 in, width = 2.3 in}
\end{minipage}

\caption{ Elastic energy density profiles in 
$2$-sheet with two sharp bends pictured in Fig~\ref{fig:2sheet-embed}.
The sheets
were $X=200$ lattice sites wide. 
The elastic thickness of each was $h = 2 \times 10^{-4} X$. 
In each plot the height of the surface is 
proportional to energy density in relative units and the $x$ and $y$
coordinates are the material coordinates in the manifold. 
The same energy units are used in all four plots.
Plots (a) and (b) are for a $3$-dimensional embedding, plots 
(c) and (d) are from a sheet embedded in $4$ dimensions.
Plots (a) and (c) are the bending energies in the
$2$-sheets, plots (b) and (d) are the stretching energies.
}
\label{fig:2sheet-enmap}
\end{figure}

\begin{figure}[tbp!]
\center
\small
\raisebox{1.9 in}{(a)}
\begin{minipage}[t]{3.0 in}
\centering
\epsfig{file=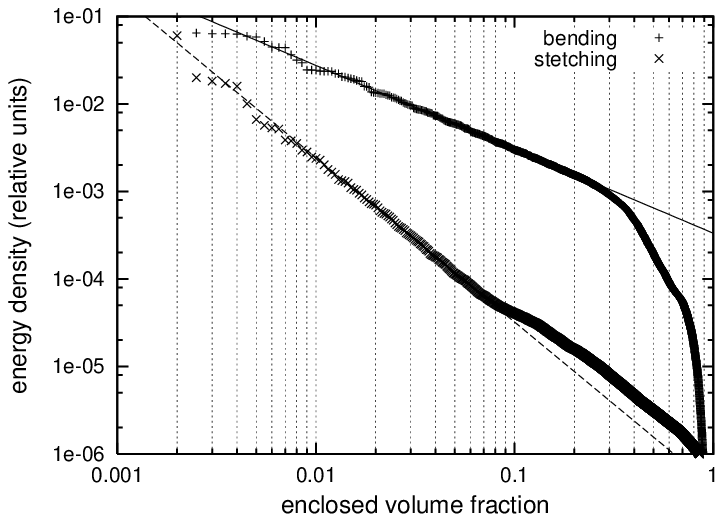, height=2.1 in, width = 3.0 in}
\end{minipage}
\hfill
\raisebox{1.9 in}{(b)}
\begin{minipage}[t]{3.0 in}
\centering
\epsfig{file=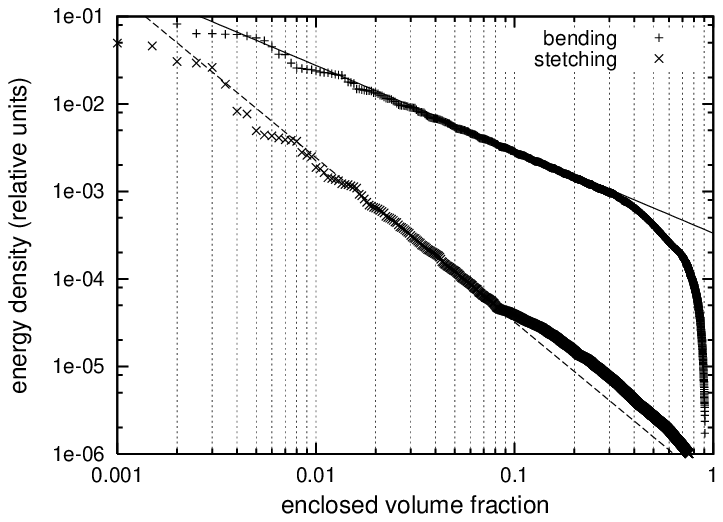, height=2.1 in, width = 3.0 in}
\end{minipage}

\caption{Energy density plots for elastic half-cubes with
single line disclinations.
The rectangular solids
were $X=40$ lattice sites across 
perpendicular to the face
with the disclination and $80$ sites wide in the other directions.
They had elastic 
thickness $h = 2.5 \times 10^{-4} X$.  
In each graph the $+$ symbols denote 
bending energy while the $\times$ symbols denote stretching energy.  
Energies are expressed in arbitrary units.  
Horizontal axes are volume fraction $\Phi$.  Graph (a) shows 
local stretching
energy density $\calL_s$ and bending energy density $\calL_b$
versus volume fraction $\Phi$ at or above this energy density
from an embedding in 4 dimensions.  
Graph (b) shows the same quantities from 
an embedding in 5 dimensions.  
In both graphs the straight lines are power law fits to the 
bending and stretching energy densities
in the region 
between $2.0 \%$ and $10 \%$ volume fraction.
In (a), the solid line is a fit to the 
bending energy density,
with scaling exponent $-0.95$, and
the dashed line is a fit to the 
stretching energy density,
with scaling exponent $-1.87$.
In (b), the solid line is a fit to the 
bending energy density, 
with scaling exponent $-0.95$, and
the dashed line is a fit to the 
stretching energy density,
with scaling exponent $-1.87$.}
\label{fig:oned_hist}
\end{figure}

\subsection{$2$-Sheet: Two Sharp Bends}
\label{sec:twosheettwodiscl}

The behavior of $2$-sheets with two disclinations in $3$-dimensional
space has been studied extensively~\cite{Alex}, and our simulations of
this geometry reproduced familiar results. However, we found for a
variety of material thicknesses and disclination geometries the behavior
of the same $2$-sheets embedded in $4$ dimensions was remarkably
different. The data presented here is for a sheet geometry which 
closely related to imposed disclinations and
which displayed the sheet's behavior particularly well.
Instead of creating a disclination like that in
Fig.~\ref{fig:makedisclin} (a), we 
fold opposite edges of the sheet and attach them to rigid
frames with sharp bends at their centers. 
Each frame keeps the edge straight
with a $90^o$ angle at its center point. The frames are
free to translate or rotate in the embedding
space. This boundary condition is close 
to the conditions used to create ``minimal'' ridges in~\cite{Alex}.
In that work Lobkovsky argued that the configuration of the sheet
around a bending point on the edge will be much like that around a
vertex. We found that the quantitative behavior of this boundary
condition was consistent with that of imposed disclinations, but it
allowed for more flexibility. The equilibrium embeddings of 
sheets with this geometry are pictured in Fig.~\ref{fig:2sheet-embed},
Fig.~\ref{fig:2sheet-hist} presents plots of energy density
versus area for $3$ and $4$ dimensional
embeddings, and Fig.~\ref{fig:2sheet-enmap} plots
local bending and stretching energy densities in the sheets' coordinate
systems. 

It is immediately evident from Figs.~\ref{fig:2sheet-enmap} that the
stretching energy density in the region between the two sharp bends is
greatly diminished in the $4$-dimensional embedding compared to the
$3$-dimensional embedding. For the latter embedding, 
the line of high stretching energy density in
Fig.~\ref{fig:2sheet-enmap}(b) marks the presence of the stretching
ridge.
However, there is no such stretching line in  the $4$-dimensional embedding
energy map plotted in Fig.~\ref{fig:2sheet-enmap}(d), even though
Fig.~\ref{fig:2sheet-embed}(b) shows that there is still a folding line
between the sharp bends in $4$ dimensions. The energy plot 
in Fig.~\ref{fig:2sheet-hist}(a), for
$3$-dimensional embedding, shows the
parallel scaling of bending and stretching energies which is indicative
of a ridge, but the energy plot in
Fig.~\ref{fig:2sheet-hist}(b), for $4$ dimensions,
is more suggestive of cone-like scaling,
since the stretching energy falls twice as fast as bending energy away
from the sharp bends.

Examining the embedding coordinates of the manifold in $4$ dimensions, 
we found that the sheet
mainly occupied only $3$ on the $4$ available spatial dimensions.
Fig.~\ref{fig:2sheet-embed}(c) plots the value of the embedding coordinate
with the lowest moment of inertia
(the moment for the entire manifold in this direction is four orders
less than that in other directions, in a frame where the inertia tensor
is diagonal). We believe
this slight bubbling into the extra dimension acts as a sink for
compressive stress along the line connecting the sharp bends.
Since this deviation is so small, one
of the two normals to the manifold lies mostly in this direction over
the entire surface. 
The curvature shown in Fig.~\ref{fig:2sheet-embed}(c) 
is small compared to the major component of
curvature across the folding line, and is nearly orthogonal to it, so
it has little effect on the total bending energy.
Yet, in the thin sheets we simulate,
the resulting changes in the strain field affect
the stretching energy enormously.

The bubbling discussed above is evidence of an interaction
between the sharp bends, 
since such a configuration is not seen for
isolated disclinations or vertices and must be energetically less
favorable than perfectly straight cone generators. If the sharp
bends
interact in a way that depends on the distance between them relative 
to the elastic thickness, then
there might be some analog of a higher
dimensional ridge between them, with much weaker stretching energy.
We did not see ridge-like 
parallel scaling in Fig.~\ref{fig:2sheet-hist}(b), but it
is possible that the strain in this kind of ridge
is so weak, and the virial ratio between bending and stretching is 
correspondingly so high, that 
the systems we simulated were dominated by a cone-like
configuration near the
sharp bends and not by the ridge between them. If this is the case,
our stretching energy graph shows only
the initial energy fall-off away from the sharp bends
and never reaches the
energy density value at which parallel scaling would commence. 
We can use the graph to 
put a lower limit on any possible virial relation by noting that
cone-like scaling continues to at least $2 \%$ volume fraction, at which
point the ratio between bending and stretch energy densities is $\approx
70$. Thus, if the bending and stretching energies do scale with elastic
thickness, they should satisfy $E_b > 70 \times E_s$. 

Following the derivation presented in Appendix \ref{sec:virialderivation}, 
we can 
use the virial relation to
put limits on the scaling exponents for the typical curvatures and strains
on the ridge. For the above virial ratio, 
in the $h \rightarrow 0$ limit the typical
mid-ridge curvature would increase more slowly than $(X/h)^{1/35}$ and the 
typical ridge strain would fall faster than $(h/X)^{34/35}$, where $h$ 
is the elastic thickness and $X$ is the length of the folding line.
To test our scaling hypothesis we probed the deviation
into the normal direction shown in 
Fig.~\ref{fig:2sheet-embed}(c). We estimate that the 
inverse square of the height of these bumps is 
proportional to the residual Gaussian 
curvature and therefore the strain along the ridge.
In simulations of the same system at several different thicknesses, 
spanning two orders of magnitude,
we could not discern a consistent change in 
the peak-to-peak height along this second normal
(see Fig.~\ref{fig:2sheet-embed} (d)).
Since our ridge scaling arguments tell us we should see clear scaling 
of this peak-to-peak height with thickness, we conclude that 
either we are 
not close enough to the thin limit for any potential scaling behavior 
to be evident, or the equilibrium configuration is really a higher
dimensional variation of simple cone scaling, which is truly independent
of elastic thickness and fold length. 
These tests were run at $(h/X)$ ratios from 
$10^{-3}$ to $10^{-5}$, the entire range of thicknesses our 
simulations can handle and a region where $2$-sheets in $3$ dimensions 
show very clear thickness scaling. It is clearly beyond our 
computational capabilities to resolve this potential scaling behavior.

\subsection{$3$-Sheet: Single Disclination}
\label{sec:onediscl}

To verify our numerical predictions for the cone, we simulate
an elastic half-cube with a single line disclination on one face. We use
a $40 \times 80 \times 80$ unit lattice. The minimum energy embedding is
a virtually identical cone in all the planes perpendicular to the 
line disclination. The radius of the cone ranges from $40$ to $\sqrt{2}
\times 40$ lattice units. Fig.~\ref{fig:oned_hist} show the
scaling of bending and stretching energy densities away from the
disclination for both $4$ and $5$-dimensional embeddings. In both cases
the scaling exponents are very close to the theoretical values of $-1$
for bending and $-2$ for stretching for a 
cone with $2$-dimensional symmetry. We are quite satisfied that the
elastic lattice can accurately represent the cone around a
single disclination.

\begin{figure}[tbp!]
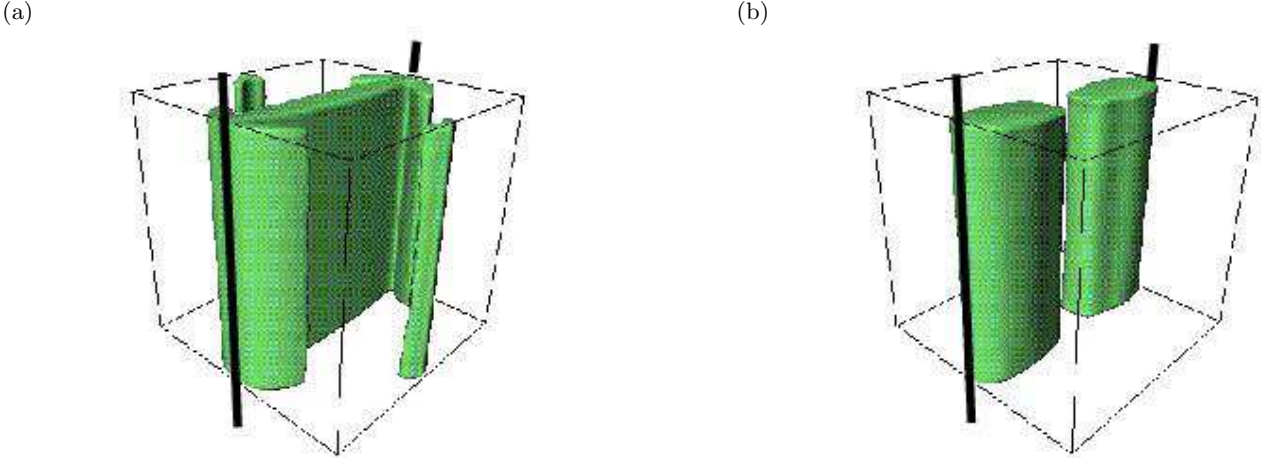

\center
\small
\raisebox{2.3 in}{(a)}
\begin{minipage}[t]{3.0 in}
\centering
\epsfig{file=figure11a.eps2, height=2.5 in, width = 2.5 in}
\end{minipage}
\hfill
\raisebox{2.3 in}{(b)}
\begin{minipage}[t]{3.0 in}
\centering
\epsfig{file=figure11b.eps2, height=2.5 in, width = 2.5 in}
\end{minipage}

\caption{ Energy condensation map for cubes with parallel disclinations. 
The cubes
were $X=80$ lattice sites wide and had elastic 
thickness $h = 2 \times 10^{-4} X$. 
Image (a) shows a surface of constant bending
energy density in the material coordinate system for a cube embedded in
$4$ dimensions. 
The surface encloses $\approx 10.0\%$ volume
fraction and shows energy condensation along the ridge which 
spans the gap between the disclinations. 
Image (b) shows the same surface for a cube embedded in
$5$ dimensions. 
The wireframes represents the edges of the cubes' material 
coordinates. Heavy lines mark the locations of disclinations.
}
\label{fig:parad_ede}
\end{figure}

\subsection{$3$-Sheet: Parallel Disclinations}
\label{sec:pdiscl}
 
Apart from boundary conditions, the cube with parallel disclinations 
has a natural symmetry along the direction of the disclinations. We
found that for all initial conditions tested, energy minimization 
resulted in a final configuration which showed this same 
symmetry (see Fig.~\ref{fig:parad_ede}). 
The manifold has no strain
or curvature in the direction parallel to the disclinations, and very similar
configurations for all planes perpendicular to this direction. In principle,
for embedding in $d$ dimensions, the configuration in each of the
perpendicular planes is identical to that which we would expect for an
elastic $2$-sheet with the same thickness to length ratio embedded in
$d-1$ dimensions. In practice, we find that the extra material dimension
adds an additional stiffness against fine scale crumpling which
often confuses similar simulations in $2$-sheets.

\begin{figure}[tbp!]
\center
\small
\raisebox{1.9 in}{(a)}
\begin{minipage}[t]{3.0 in}
\centering
\epsfig{file=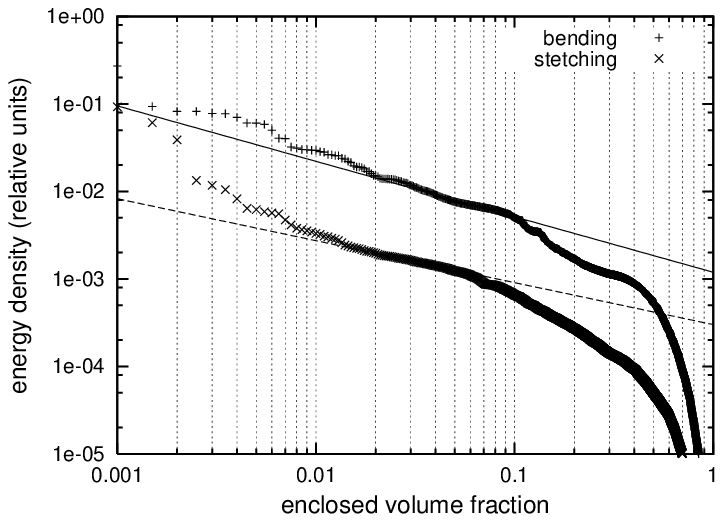, height=2.1 in, width = 3.0 in}
\end{minipage}
\hfill
\raisebox{1.9 in}{(b)}
\begin{minipage}[t]{3.0 in}
\centering
\epsfig{file=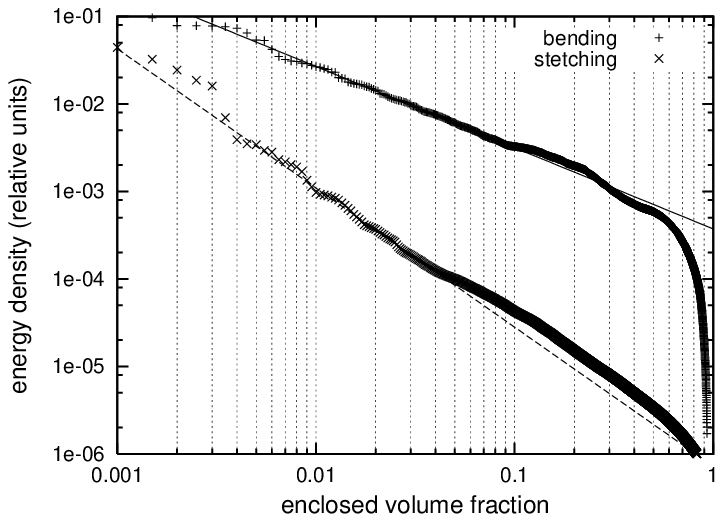, height=2.1 in, width = 3.0 in}
\end{minipage}

\caption{Energy density plots for an elastic cube with
parallel disclinations 
The cubes
were $X=80$ lattice sites wide and had elastic 
thickness $h = 2 \times 10^{-4} X$.  
In each graph the $+$ symbols denote 
bending energy while the $\times$ symbols denote stretching energy.  
Energies are expressed in arbitrary units.  
Horizontal axes are volume fraction $\Phi$.  Graph (a) shows 
local stretching
energy density $\calL_s$ and bending energy density $\calL_b$
versus volume fraction $\Phi$ at or above this energy density
from an embedding in 4 dimensions.  
Graph (b) shows the same quantities from 
an embedding in 5 dimensions.  
In both graphs the straight lines are power law fits to the 
bending and stretching energy densities.
In graph (a) the bending energy fit is to the region 
between $3.0 \%$ and $8.0 \%$ volume fraction
and the stretching energy fit is to the region 
between $2.0 \%$ and $5.0 \%$ volume fraction.
In graph (b) the fits are to the region 
between $1.0 \%$ and $4.0 \%$ volume fraction.
In (a), the solid line is a fit to the 
bending energy density, with scaling exponent $-0.63$, and
the dashed line is a fit to the 
stretching energy density, with scaling exponent $-0.48$.
In (b), the solid line is a fit to the 
bending energy density, with scaling exponent $-0.93$, and
the dashed line is a fit to the 
stretching energy density, with scaling exponent $-1.59$.}
\label{fig:parad_hist}
\end{figure}

For $4$-dimensional embedding the equilibrium configuration
is a ``stack of ridges,'' 
which shows the same energy scaling as a
ridge in $3$ dimensions.
Fig.~\ref{fig:parad_hist} (a) is a plot of energy density versus volume
for a $3$-cube with parallel disclinations in $4$ dimensions.
Within the high energy region encompassing 
$\approx 2 - 8 \%$ volume fraction, 
the ratio of bending energy density to stretching energy density at a
given volume fraction is $\approx 6.2$. This number is consistent
with the theoretical energy ratio of $5$. In this volume range
the plots also confirm the lack of
condensation {\em along} the ridge of both 
bending and stretching energies 
as well as the identical scaling of these energies.
For the entire region up to $\approx 30 \%$ volume fraction
the bending and
stretching energy have roughly the same dependence on volume,
though they don't fit a clean scaling exponent for any extended region.
The sharp drop-off of the
dominant energy above $30 \%$ volume fraction shows the significant
condensation of energy around the ridge structure.

In contrast to the ridge-scaling in $4$ dimensions,
the energy scaling behavior 
of the sheet embedded in $5$ dimensions 
appears cone-like. The stretching scaling
exponent of $-1.59$ indicates that the stretching energy is condensed at
the vertices, while the bending exponent of $-0.93$ is consistent with
$-1$, the predicted
bending scaling of isolated line disclinations. The scaling data, as
well as the lack of a strong ridge region in Fig~\ref{fig:parad_ede}(b),
indicate that the scaling around each disclination is not strongly
influenced by interaction between the two disclinations.
The scaling resembles that around the
isolated disclination reported in Section~\ref{sec:onediscl}.

\begin{figure}[tbp!]
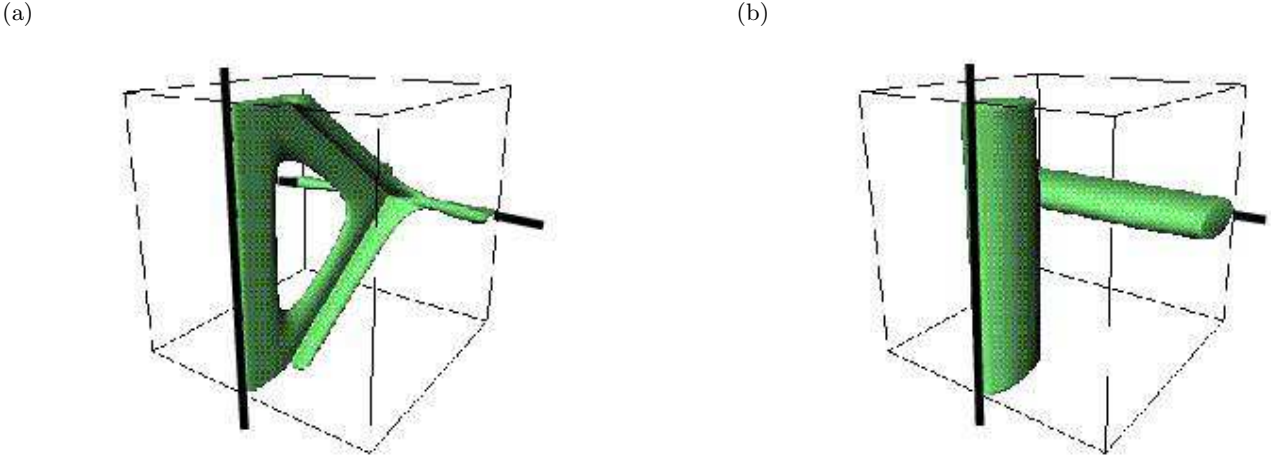

\center
\small
\raisebox{2.3 in}{(a)}
\begin{minipage}[t]{3.0 in}
\centering
\epsfig{file=figure13a.eps2, height=2.5 in, width = 2.5 in}
\end{minipage}
\hfill
\raisebox{2.3 in}{(b)}
\begin{minipage}[t]{3.0 in}
\centering
\epsfig{file=figure13b.eps2, height=2.5 in, width = 2.5 in}
\end{minipage}

\caption{ Energy condensation map for cubes with perpendicular
disclinations.  The cubes
were $X=60$ lattice sites across, $90$ lattice units wide in the
directions parallel to the disclinations,
and had elastic 
thicknesses $h = 6 \times 10^{-4} X$ (a) and 
$h = 3 \times 10^{-4} X$ (b). 
Image (a) shows a surface of constant bending
energy density in the material coordinate system for a cube
embedded in $4$~dimensions. 
Image (b) shows a surface of constant bending
energy density for a cube embedded in $5$ dimensions.
The energy density value encloses
$\approx 3\%$ volume fraction in image (a) and shows the spontaneous 
vertex lines which arise between disclinations. In image (b) the 
equal energy density surface encloses $\approx 10\%$ volume 
fraction and shows the growth of the high energy region 
around the disclinations without additional vertices evident 
between them.
The wireframes represent the edges of the cubes' material 
coordinates. Heavy lines mark the locations of disclinations.
}
\label{fig:3disclin45d_ede}
\end{figure}

\begin{figure}[tbp!]
\center
\small
\raisebox{1.9 in}{(a)}
\begin{minipage}[t]{3.0 in}
\centering
\epsfig{file=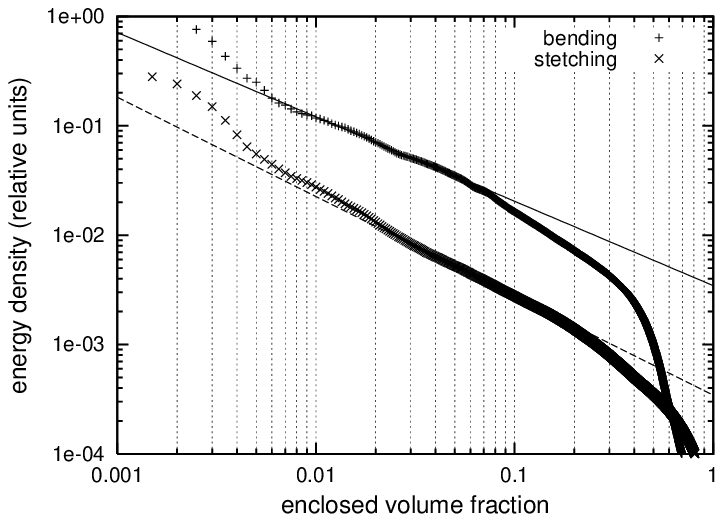, height=2.1 in, width = 3.0 in}
\end{minipage}
\hfill
\raisebox{1.9 in}{(b)}
\begin{minipage}[t]{3.0 in}
\centering
\epsfig{file=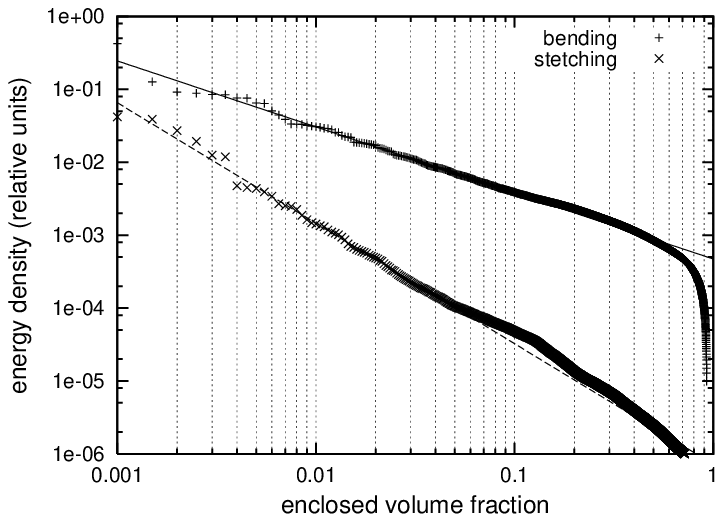, height=2.1 in, width = 3.0 in}
\end{minipage}

\caption{Energy density plots for the elastic cubes with
non-parallel disclinations in Fig~\ref{fig:3disclin45d_ede}.  The 
cubes
were $X=60$ lattice sites across, $90$ lattice units wide in the
directions parallel to the disclinations, and had elastic 
thicknesses $h = 6 \times 10^{-4} X$ (a) and 
$h = 3 \times 10^{-4} X$ (b).  In each graph the $+$ symbols denote 
bending energy while the $\times$ symbols denote stretching energy.  
Energies are expressed in arbitrary units.  
Horizontal axes are volume fraction $\Phi$.  Graph (a) shows 
local stretching
energy density $\calL_s$ and bending energy density $\calL_b$
versus volume fraction $\Phi$ at or above this energy density
from an embedding in 4 dimensions.  Graph (b) shows the same quantities from 
an embedding in 5 dimensions.  
In (a), the solid line is a power law fit to the 
bending energy density in the region between $1.0 \%$ and $5.0 \%$ 
volume fraction, with scaling exponent $-0.77$, and
the dashed line is a power law fit to the 
stretching energy density in the region between $3.0 \%$ and $10 \%$ 
volume fraction, with scaling exponent $-0.91$.
In (b), the solid line is a power law fit to the 
bending energy density in the region between $1.0 \%$ and $10.0 \%$ 
volume fraction, with scaling exponent $-0.90$, and
the dashed line is a power law fit to the 
stretching energy density in the region between $1.0 \%$ and $5 \%$ 
volume fraction, with scaling exponent $-1.65$.}
\label{fig:3disclin45d_hist}
\end{figure}

\subsection{$3$-Sheet: Perpendicular Disclinations}
\label{sec:perpdiscl}

A typical spatial energy distribution after
energy minimization for a cube with perpendicular
disclinations embedded in $d=4$ is
shown in Fig.~\ref{fig:3disclin45d_ede}(a). 
A common feature of all the
cubes embedded in $4$ dimensions
is the spontaneous appearance of additional
line-like vertex structures. Spanning the volume between vertices
and disclinations are $2$-dimensional ridges.
This result is consistent with the $d=4$ geometrical 
confinement simulations
detailed in~\cite{eric}. 
Fig.~\ref{fig:3disclin45d_hist}(a) shows the decay of local energy density 
with volume away from ridges and
vertices for an elastic cube with
non-parallel disclinations in $4$ dimensions.
The highest-energy regions correspond to the imposed disclinations
and spontaneous vertex network. 
Lower energy regions correspond to ridges.
The region between $1\% $ and
$\approx 5 \%$ volume fraction shows smooth scaling of bending 
and stretching energy densities
with volume in a region dominated by the high energy part of
ridge structures (where they join at vertices). 
In this region
the bending energy density scales with
exponent $-0.77$ and the stretching with
exponent $-0.91$.
The values of the scaling exponents in this region are reasonably
close to each other and to the theoretical scaling of $-4/5$ derived in
Section~\ref{sec:scaling}.
The scaling is clearly distinct from that of a cone,
where the stretching energy density is expected to fall faster than 
$\Phi^{-1}$.

\begin{figure}[tbp!]
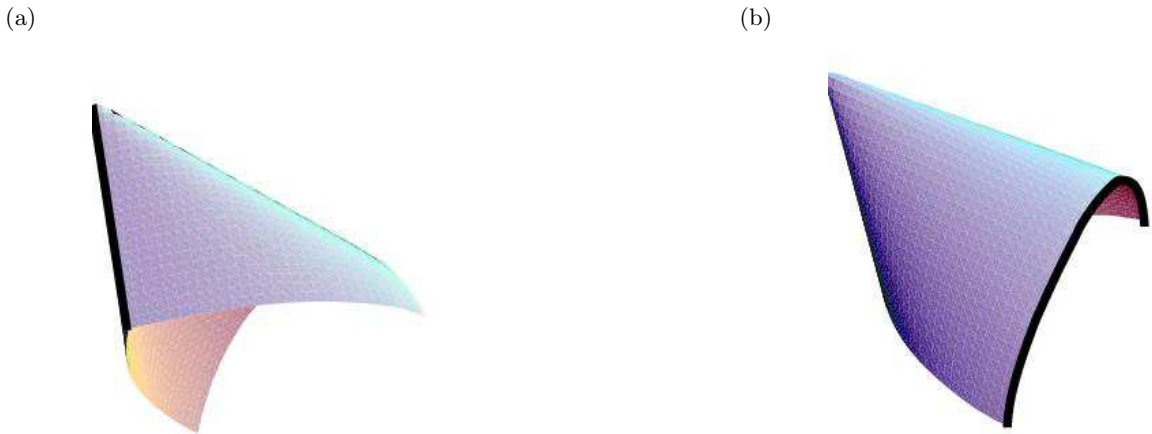

\center
\small
\raisebox{2.3 in}{(a)}
\begin{minipage}[t]{3.0 in}
\centering
\epsfig{file=figure15a.eps2, height=2.5 in, width = 2.5 in}
\end{minipage}
\hfill
\raisebox{2.3 in}{(b)}
\begin{minipage}[t]{3.0 in}
\centering
\epsfig{file=figure15b.eps2, height=2.5 in, width = 2.5 in}
\end{minipage}

\caption{Two views of a
$3$-dimensional projection of the embedding coordinates for a
$2$-slice from the material coordinates of a cube with non-parallel
disclinations embedded in $5$ dimensions. The material coordinate slice
contains one disclination along the edge marked with the heavy line in (a).
It is perpendicular to the
other disclination, which is marked with a heavy line in (b).
The views were chosen to show the cone-like
geometry about the disclination which the plane intersects at a point.
}
\label{fig:5slice-cone}
\end{figure}

The behavior of the same cubes placed in $d=5$ was quite different 
(see Fig.~\ref{fig:3disclin45d_ede}~(b)). For this embedding 
there is no spontaneous
ridge-vertex network between the imposed disclinations. The difference
in structure is reflected in the energy density plot,
Fig.~\ref{fig:3disclin45d_hist}(b). In the volume fraction 
which typically encompasses high energy structures apart
from vertices and disclinations, the bending energy scales with volume
with an exponent of $-0.90$ while the stretching energy scales with an
exponent of $-1.65$. These numbers indicate the dominance of
conical scaling near the vertices.
They are similar to the 
scaling exponents for the cube with parallel disclinations in $5$
dimensions. 
Fig.\ref{fig:5slice-cone}
shows that the embedding appears to be
locally conical around a disclination, without any evidence of
folds. We surmise that the sheet has relaxed to a 
configuration where the cones around each line vertex 
interpenetrate without interacting strongly. 
This result demonstrates that in higher dimensional crumpling, multiple
vertices can exist in a sheet without requiring folds, for
some geometries.

For this simulated geometry and several of the following, we present 
data for thicker sheets in $4$ dimensions than in $5$. This is 
because the ridges found in $4$ dimensions become very sharp as $h$ 
gets smaller, and we typically present the thinnest data which does 
not show signs of finite lattice effects (like those seen below in
Fig~\ref{fig:3tin45d_hist}(a)). For $5$-dimensional embeddings, 
decreasing the thickness typically shifts all the 
stretching energy densities upward and thereby
extends the volume of cone 
scaling visible before the stretching energy density fades 
to background levels.
We thus chose to present data from thinner sheets for $5$ dimensions. 
We found that ridge scaling becomes more distinct as the 
sheet becomes thinner, so 
the use of thinner sheets in $5$ dimensions only strengthens our claim that 
there is no evident ridge structure in $5$ dimensions.


\begin{figure}[tbp!]
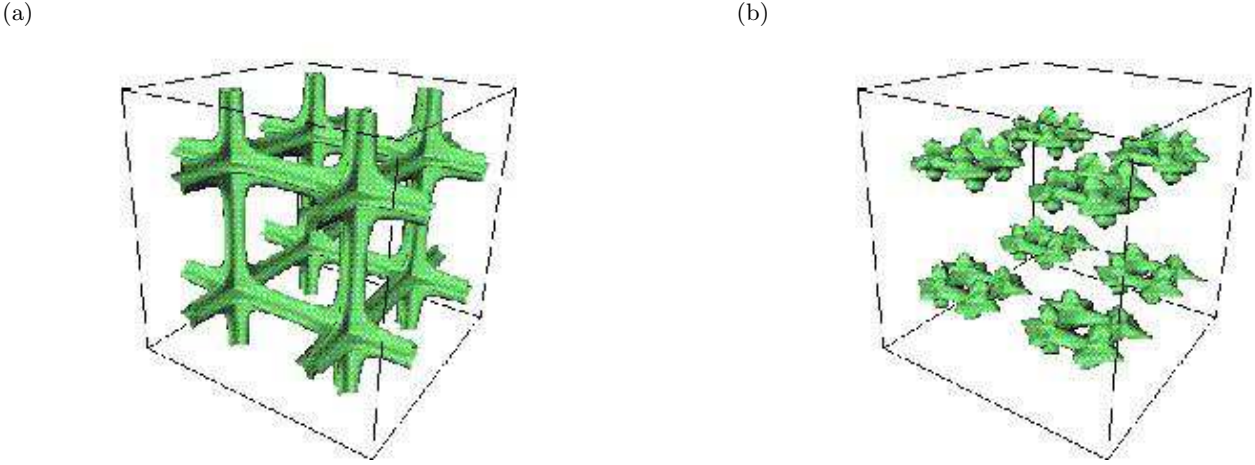

\center
\small
\raisebox{2.3 in}{(a)}
\begin{minipage}[t]{3.0 in}
\centering
\epsfig{file=figure16a.eps2, height=2.5 in, width = 2.5 in}
\end{minipage}
\hfill
\raisebox{2.3 in}{(b)}
\begin{minipage}[t]{3.0 in}
\centering
\epsfig{file=figure16b.eps2, height=2.5 in, width = 2.5 in}
\end{minipage}

\caption{Energy condensation maps for $3$-tori. The tori
were made from cubes of width $X=80$ lattice sites and with elastic 
thicknesses $h = 1 \times 10^{-3} X$ (a) and 
$h = 2.5 \times 10^{-4} X$ (b). 
Image (a) shows a surface of constant bending
energy density in the material coordinate system for a $3$-torus
embedded in $4$ dimensions. In this simulation the initial conditions 
were chosen to favor a symmetric relaxed state.
Image (b) shows a surface of constant bending
energy density for a $3$-torus embedded in $5$ dimensions.
The surfaces encloses $\approx 2.5\%$ volume
fraction in (a) and  $\approx 2.8\%$ volume
fraction in (b).
The wireframes represent the edges of the cubes' material 
coordinates.}
\label{fig:3tin45d_edens}
\end{figure}

The analog of ridges in the $3$-torus in $d=4$ are again 
planes of high elastic energy.
These ridge structures meet in vertex-lines of very high 
elastic energy.
Fig~\ref{fig:3tin45d_hist}(a) shows the decay of local energy density 
with volume away from ridges and
vertices for a $3$-torus in $4$ dimensions.
We fit a power law
to the region which we
identify as parts of ridges near the vertex structures.
In this region
the energy densities scale with
an exponent of $-0.87$ for bending energy and $-0.88$ for
stretching. These values are consistent with the theoretical ridge scaling
exponent of $-4/5$ derived in Section~\ref{sec:scaling}.
Within the ridge scaling volume
the ratio of bending energy density to stretching
energy density is $8.5$,
within a factor of two of
the known value, $5$, for $2$-sheets in $3$-dimensional crumpling.

\begin{figure}[tbp!]
\center
\small
\raisebox{1.9 in}{(a)}
\begin{minipage}[t]{3.0 in}
\centering
\epsfig{file=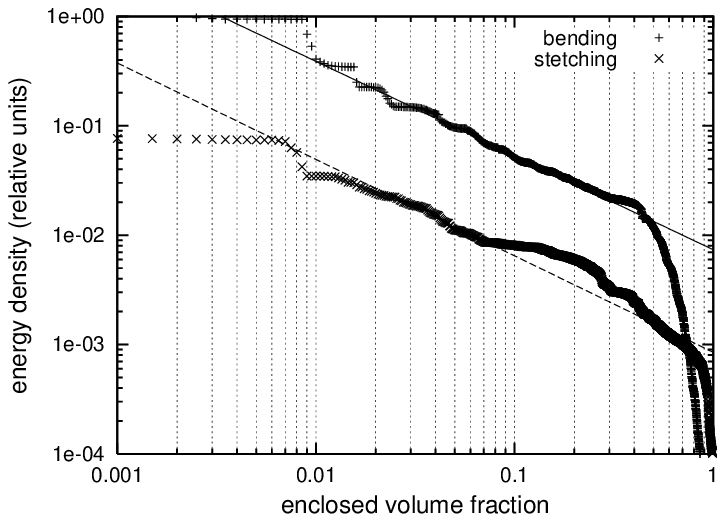, height=2.1 in, width = 3.0 in}
\end{minipage}
\hfill
\raisebox{1.9 in}{(b)}
\begin{minipage}[t]{3.0 in}
\centering
\epsfig{file=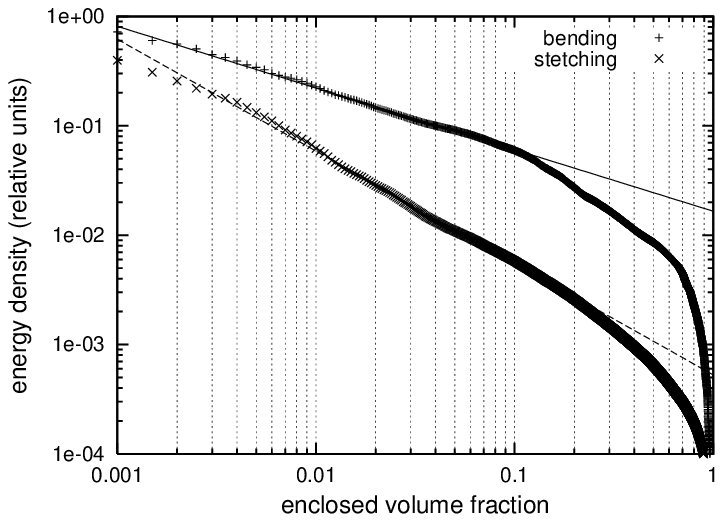, height=2.1 in, width = 3.0 in}
\end{minipage}

\caption{Energy density plots for the $3$-tori in 
Fig~\ref{fig:3tin45d_edens}.  The tori
were made from cubes of width $X=80$ lattice sites and with elastic 
thicknesses $h = 1 \times 10^{-3} X$ (a) and 
$h = 2.5 \times 10^{-4} X$ (b).  
In each graph the $+$ symbols denote 
bending energy while the $\times$ symbols denote stretching energy.  
Energies are expressed in arbitrary units.  
Horizontal axes are volume fraction $\Phi$.  Graph (a) shows 
local stretching
energy density $\calL_s$ and bending energy density $\calL_b$
versus volume fraction $\Phi$ at or above this energy density
from an embedding in 4 dimensions.  
The several plateaus in the high energy part of the plot are an 
artifact of the discrete lattice. They reflect the 
nearly identical geometry of the points on and adjacent to the 
vertices, which make up a measureable fraction of the manifold volume. 
Graph (b) shows the same quantities as in (a) for 
an embedding in 5 dimensions.  
In (a), the solid line is a power law fit to the 
bending energy density in the region between $2.0 \%$ and $10 \%$ 
volume fraction, with scaling exponent $-0.87$, and
the dashed line is a power law fit to the 
stretching energy density in the region between $2.0 \%$ and $6.0 \%$ 
volume fraction, with scaling exponent $-0.88$.
In (b), the solid line is a power law fit to the 
bending energy density in the region between $0.5 \%$ and $10.0 \%$ 
volume fraction, with scaling exponent $-0.56$, and
the dashed line is a power law fit to the 
stretching energy density in the region between $0.5 \%$ and $10 \%$ 
volume fraction, with scaling exponent $-1.02$.}
\label{fig:3tin45d_hist}
\end{figure}

\section{Toroidal Connectivity}
\label{sec:torus}

Toroidal connectivity was simulated numerically by
defining the lattice displacement vector, $\ulDelta$,
such that opposite faces of our cubic array had
nearest-neighbor connections. The resulting connectivity
was everywhere isotropic, with no borders or disclinations.
Simulations were run for $3$-tori embedded in 4,5 and
6 spatial dimensions. Initial conditions were
either random or chosen to be very symmetric
or close to possible energy minima.
In all cases when $d=4$ or $5$, minimization
of the elastic energy resulted in energy condensation and the 
formation of high-energy networks. 
For $6$-dimensional embeddings
the elastic energy was many times smaller 
than in lower dimensions and was uniformly distributed over the manifold.
Figure.~\ref{fig:3tin45d_edens} compares the energy condensation
networks for $3$-tori
embedded in either $4$ or $5$ dimensions. 
Although the network is more extensive for 
$d=4$, in either dimension high-energy structures have comparable
energy densities. 
We simulated this geometry beginning from
many different initial conditions and 
the resulting final configurations showed varying degrees of asymmetry 
between the vertices.
The configuration presented in
Fig.~\ref{fig:3tin45d_edens}(a) showed the highest degree of 
symmetry of all our $4$-dimensional torus simulations -- its total elastic
energy was $\sim 25\%$ less than that of configurations which broke 
symmetry, so we believe it is most likely the true ground state of the 
system. Our energetic analysis was performed on this configuration.

The energy structures of tori embedded 
in $d=5$ were qualitatively
different from those embedded in $d=4$ (refer again to 
Fig.~\ref{fig:3tin45d_hist}).
For $d=5$, the structures corresponding to vertices appear point-like
instead of line-like. 
The majority of the total energy density is concentrated
around these point-like vertices.
Between vertices we were able to see smaller, line-like energy
concentrations of elastic energy which could correspond to ridge
structures. 
However, these regions occupied a miniscule volume in the manifold. The
predominant energy structures were  more symmetric and were centered
around vertices. 

Energy density versus volume 
is plotted in Fig~\ref{fig:3tin45d_hist}(b) 
for an 80 lattice unit $3$-torus 
embedded in $d=5$. 
Smooth energy scaling begins at about $0.5 \% $ volume fraction
and holds for up to $\approx 10 \%$ of the total volume.
Within these high energy regions the bending energy density
scales with volume with an exponent of $-0.56$, whereas stretching 
energy density dies off more quickly,  
with a scaling exponent of $-1.02$.
This is consistent with our simulations of line
disclinations in $5$ dimensions, since the stretching energy 
falls off nearly twice as fast as the bending energy.
The number of vertices present in the manifold would lead us to expect
ridges, but as we saw in Sec.\ref{sec:pdiscl},
ridges cannot be resolved by energy scaling
alone in $5$ dimensions at elastic thicknesses accessible to our 
simulations. 
The scaling exponents above are closer to the
expected exponents of $-2/3$ and $-4/3$ for conical scaling around a
point-like disclination than to any other kind of scaling behavior we 
know. The embedding is probably close to this form of cone near the
vertices. 


\begin{figure}[tbp!]
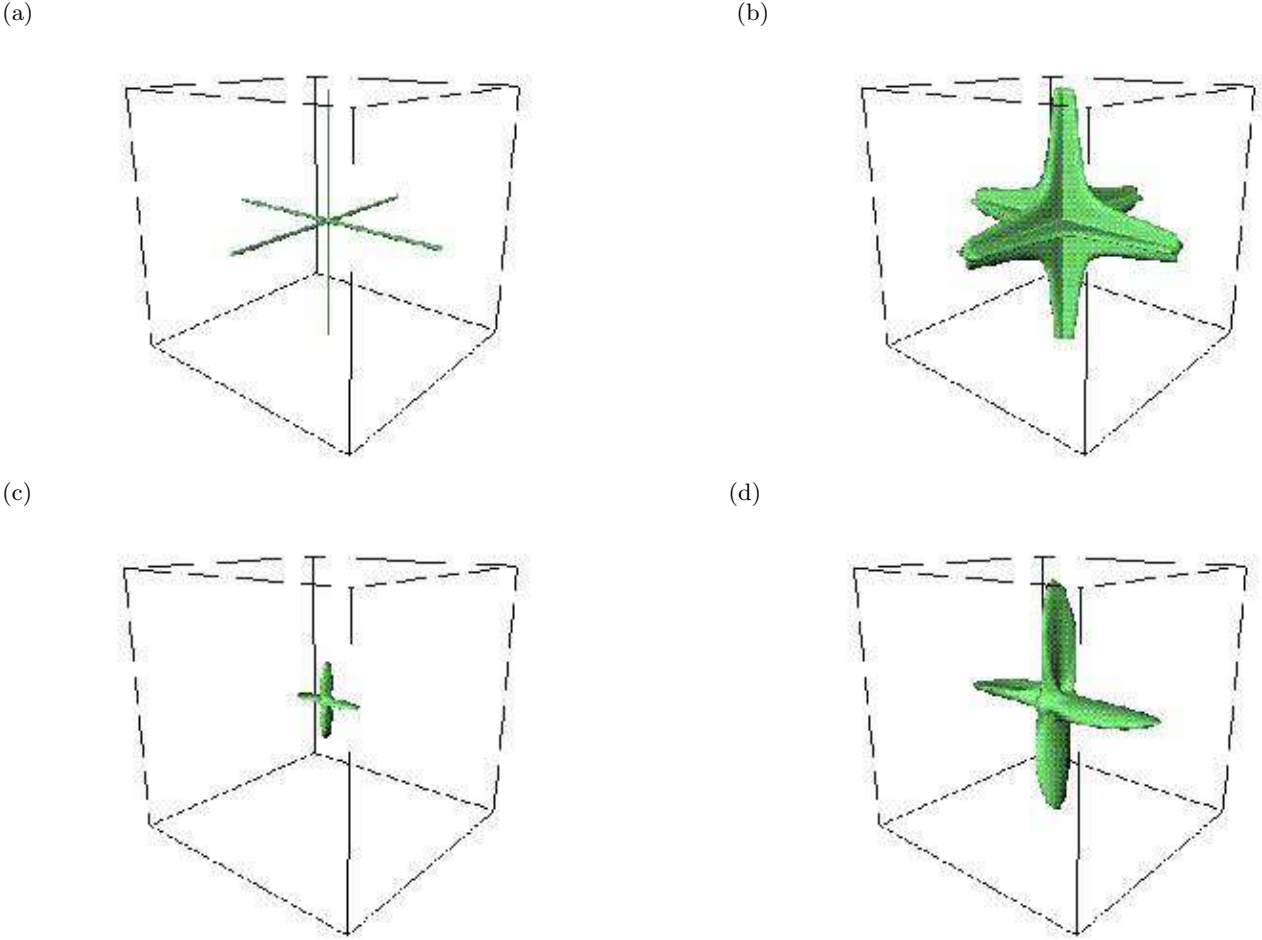

{\center
\small
\raisebox{2.3 in}{(a)}
\begin{minipage}[t]{3.0 in}
\centering
\epsfig{file=figure18a.eps2, height=2.5 in, width = 2.5 in}
\end{minipage}
\hfill
\raisebox{2.3 in}{(b)}
\begin{minipage}[t]{3.0 in}
\centering
\epsfig{file=figure18b.eps2, height=2.5 in, width = 2.5 in}
\end{minipage}

\raisebox{2.3 in}{(c)}
\begin{minipage}[t]{3.0 in}
\centering
\epsfig{file=figure18c.eps2, height=2.5 in, width = 2.5 in}
\end{minipage}
\hfill
\raisebox{2.3 in}{(d)}
\begin{minipage}[t]{3.0 in}
\centering
\epsfig{file=figure18d.eps2, height=2.5 in, width = 2.5 in}
\end{minipage}
}

\caption{Energy condensation maps for $3$-cubes with center points of
opposite faces attached. The cubes
were $X=80$ lattice sites wide and had elastic 
thickness $h = 2.5 \times 10^{-4} X$.
Images (a) and (b) show surfaces of constant bending
energy density in the material coordinate system for a $3$-cube
embedded in $4$ dimensions. In this simulation the initial conditions 
were chosen to favor a symmetric relaxed state -- many stable 
configurations show pronounced symmetry breaking in one direction.
Images (c) and (d) show surfaces of constant bending
energy density for a $3$-cube embedded in 
$5$ dimensions. The surfaces in (a) and (c)
encloses $0.1 \%$ volume fraction while those in (b) and (d) 
encloses $1 \%$.
The wireframes represent the edges of the cubes' material 
coordinates.
}
\label{fig:fold_edens}
\end{figure}

Figure~\ref{fig:fold_edens} shows the pattern of energy condensation in
$3$-cubes with attached opposite faces after energy minimization. For
$4$-dimensional embeddings the surfaces of constant energy density
enclose line-like regions which traverse the cube as seen in
Figure~\ref{fig:fold_edens}~(a) and (b). 
The high energy regions appear line-like
all the way up to the highest values of the energy density.
In contrast, the surfaces of constant energy density for $5$-dimensional
embeddings form a series of shells, as seen in
Figure~\ref{fig:fold_edens}~(c) and (d), whose typical diameters
increase with decreasing energy density value. The energy density at the
surface in Figure~\ref{fig:fold_edens}~(c), which encloses $0.1 \% $
volume fraction, is nearly an order of magnitude greater than the
energy density at the surface of Figure~\ref{fig:fold_edens}~(d), which
encloses $1 \%$ volume fraction and just touches the outside edges
of the cube. These data clearly support our assertion that point-like
vertex structures are possible in $5$-dimensional embeddings. At the same
time, they are consistent with conjectures~\cite{Alex} that the high
energy regions in $4$-dimensional embeddings are line-like. It may be
noted that the $5$-dimensional embedding is asymmetric, while the
$4$-dimensional embedding has a high degree of symmetry. We found that
the elastic manifolds always spontaneously broke symmetry
in $5$ dimensions, but the minimum energy configuration we could find
in $4$ dimensions was perfectly symmetrical.

\begin{figure}[tbp!]
{    
\center
\small
\raisebox{1.9 in}{(a)}
\begin{minipage}[t]{3.0 in}
\centering
\epsfig{file=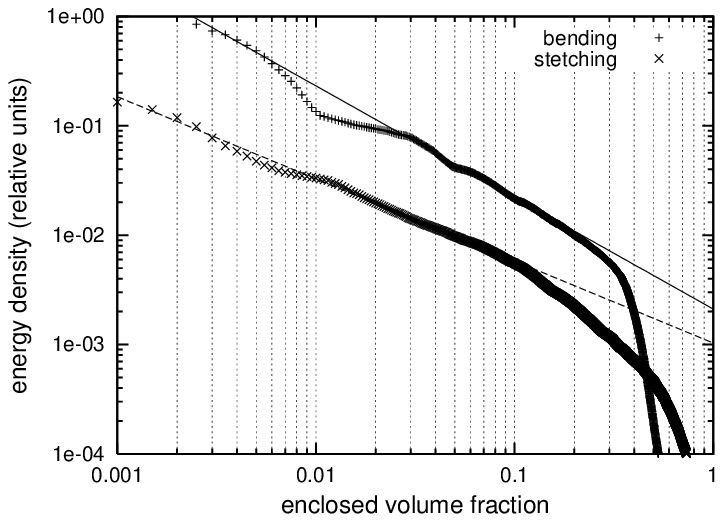, height=2.1 in, width = 3.0 in}
\end{minipage}
\hfill
\raisebox{1.9 in}{(b)}
\begin{minipage}[t]{3.0 in}
\centering
\epsfig{file=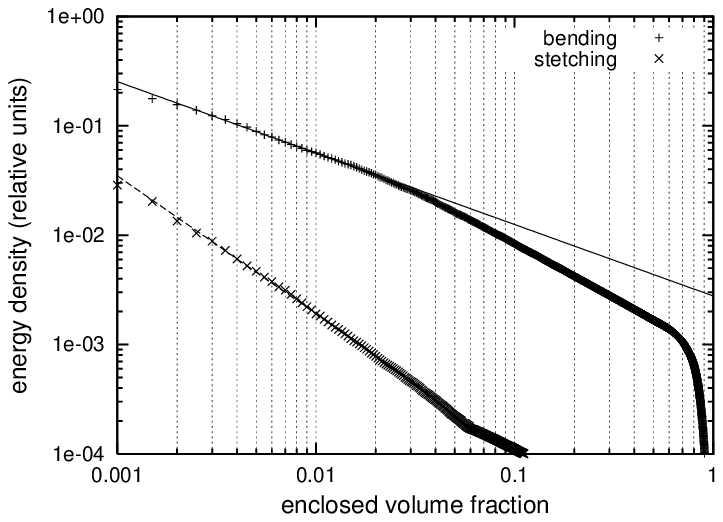, height=2.1 in, width = 3.0 in}
\end{minipage}
}

\caption{Energy density plot for the elastic cubes with
center points of
opposite faces attached pictured in Fig~\ref{fig:fold_edens}. 
The cubes
were $X=80$ lattice sites wide and had elastic 
thickness $h = .001 X$.  In each graph the $+$ symbols denote 
bending energy while the $\times$ symbols denote stretching energy.  
Energies are expressed in arbitrary units.  
Horizontal axes are volume fraction $\Phi$.  Graph (a) shows 
local stretching
energy density $\calL_s$ and bending energy density $\calL_b$
versus volume fraction $\Phi$ at or above this energy density
from an embedding in $4$ dimensions.  Graph (b) shows the same quantities
from an embedding in $5$ dimensions
In (a), the solid line is a power law fit to the 
bending energy density in the region between $3.0 \%$ and $10 \%$ 
volume fraction, with scaling exponent $-1.02$, and
the dashed line is a power law fit to the 
stretching energy density in the region between $2.0 \%$ and $9.0 \%$ 
volume fraction, with scaling exponent $-0.75$.
In (b), the solid line is a power law fit to the 
bending energy density in the region between $0.2 \%$ and $2.0 \%$ 
volume fraction, with scaling exponent $-0.65$, and
the dashed line is a power law fit to the 
stretching energy density in the region between $0.2 \%$ and $2.0 \%$ 
volume fraction, with scaling exponent $-1.26$.}
\label{fig:fold_hist}
\end{figure}

\section{Single Fold (Bow Configuration)}
\label{sec:singlefold}

In our final set of simulations we set our boundary and initial
conditions to create single, point-like vertices. Our aim was to verify
scaling predictions for vertex deformations in $4$ and $5$ dimensions
and to show clearly the existence of point-like vertex structures 
in $5$-dimensional 
embeddings. According to the relations presented in
Section~\ref{sec:theory}, a vertex is 
expected to have high Gaussian curvature. 
In more specific terms this means 
a vertex is a locus of strong curvature in at least two material directions along the same normal
vector. Therefore, to force the existence of exactly one vertex 
in $5$ dimensions (where every manifold point has two independent
normals) we searched for a minimal set of boundary conditions
which necessitated that some points in the manifold have
curvature in all three material directions.
Our most successful simulations, presented here 
for $4$ and $5$-dimensional embeddings of a $3$-cube,
had only the center points of opposite faces attached. 
This geometry caused vertices to form while leaving the majority of 
the boundary free.

Plots of energy density versus enclosed volume fraction for this
geometry are presented in Figure~\ref{fig:fold_hist}. For
$4$-dimensional 
embeddings the bending energy density does not scale with 
a simple exponent in the volume range between $1\%$ and $10\%$ volume
fraction which we associate with the high energy region of ridges. To
get a representative value of the energy drop-off we fit a power law to
the region between $3\%$ and $10\%$ volume fraction and find an exponent of 
$-1.02$. 
The stretching energy scales more cleanly in the region between 
$2\%$ and $9\%$ volume fraction, with an exponent of $-0.75$. 
The stretching energy exponent is close to the expected value of $-4/5$
and is smaller in magnitude
than $-1$, so we can safely identify the scaling as
ridge-like. When viewing the equal energy surfaces at larger volume
fractions (not shown here)
we saw a good deal of secondary structure in the ridges themselves,
which could explain the many features in the energy density dependence.

For the $5$-dimensional embedding we fit
a simple power law to the region of the graph which enclosed less than
$\approx 2 \%$ volume fraction, since above this enclosed volume
fraction the surfaces shown in Figure~\ref{fig:fold_edens}~(c) and (d)
intersect the boundary of the cube.~\cite{foot2}
In this region the
bending energy density scales with an exponent of $-0.65$ while the
stretching energy density scales with an exponent of $-1.26$. These
numbers are consistent with the theoretical $5$-dimensional cone
scaling exponent of $-2/3$ and $-4/3$ derived in 
Section~\ref{sec:scaling}.


\section{Discussion}
\label{sec:discussion}

\begin{table}

\center

\begin{tabular}{|l|c|c|c|c|c|}\hline
geometry & $m$ & $d$  & vertex dimension & scaling 
& spontaneous  assymetry  \\ \hline\hline
two sharp bends       & 2 & 3 & n/a & ridge & n/a  \\
			& 2 & 4 & n/a & cone  & n/a  \\ \hline
isolated disclinations	& 3 & 4 & n/a & cone  & n/a  \\
			& 3 & 5 & n/a & cone  & n/a  \\ \hline
$\parallel$ disclinations 	& 3 & 4 & n/a & ridge & n/a  \\
			& 3 & 5 & n/a & cone  & n/a  \\ \hline
$\perp$ disclinations 	& 3 & 4 & 1   & ridge & n/a  \\
			& 3 & 5 & n/a & cone  & n/a  \\ \hline
torus			& 3 & 4 & 1   & ridge & no  \\
			& 3 & 5 & 0   & cone  & yes   \\ \hline
bow			& 3 & 4 & 1   & ridge & no  \\
			& 3 & 5 & 0   & cone  & yes   \\ \hline

\end{tabular}

\caption{Summary of simulational results. The first three columns list the
geometry of the simulated sheet and the spatial and
embedding dimensions. 
The next column list the
dimension of any {\it spontaneous} vertex structures. The next column
tells whether the observed scaling was cone-like or ridge-like. 
The final
column tells, for the cases where the boundary conditions
did not explicitly break the $3$-dimensional 
symmetry of the cube, whether the energy minimum was a
symmetric state in the manifold coordinates. The six
manifold geometries, in the order presented here, are: two
point disclinations in a square sheet, a single
line-disclination at one cube face, parallel line disclinations on
opposite cube faces, perpendicular line disclinations on
opposite cube faces, toroidal connectivity of a $3$-dimensional
manifold, and the
attachments of the center points of opposite cube faces to each other.
}
\label{table:simsum}
\end{table}

In the numerical simulations
reported in this paper we have investigated the behavior of $2$ and
$3$-dimensional manifolds embedded in dimensions $3$ and greater,
subject to a variety of boundary conditions which cause crumpling. 
The results of our simulations, which are summarized in
Table~\ref{table:simsum}, show a consistent dependence of the
crumpling response on $d-m$, 
the difference between the dimensionality of the embedding space and
the that of the sheet. 
The behavior summarized in
Table~\ref{table:simsum}  can be
described by the following general principles:

\begin{enumerate}

\item For all the boundary conditions we consider, the dimensionality
of spontaneous vertices 
 $\calD$ is $2d -m - 1$,
suggesting that the vertex
dimensionality is always given by the {\em lower} bound from the arguments
in Section~\ref{sec:isometric} that yield $\mathrm{dim}(\calD)
\geq 2m -d -1$.

\item If $d = m+1$, the details of the curvature defect set $\calK$ 
determine the nature of the energy condensation. For $\calD =
\calK$ (no folding lines) we have the cone scaling 
discussed in Section~\ref{sec:scaling}, and for ${\cal
K} - \calD \neq \varnothing$ (folding lines present) 
we find ridge scaling.

\item If $d \geq m+2$ we always find cone scaling 
if $\calK \neq \varnothing$.

\end{enumerate}

We believe that these principles are true in general. Consequently,
they give explicit predictions for the behavior of elastic sheets in
higher dimensions. 

\subsection{Effect of Embedding Dimension on Defect Dimension}

Our data consistently supports the arguments presented in
Section~\ref{sec:isometric} that for an $m$-dimensional manifold in
$d$-dimensional space with $d < 2m$, \tw
the dimensionality of the set $\calD$ of vertex
singularities will be greater 
than or equal to $2m-d-1$. \tw 
In fact, we found
that all spontaneous vertex structures had dimensionality $2m-d-1$
identically.  For $5$-dimensional embeddings of $3$-sheets in which we
did not explicitly make line-like disclinations, the manifolds were
able to relax to configurations wherein $\calD$ was small and
point-like. In contrast, when the embedding space was $4$-dimensional
for the same manifolds and boundary conditions, $\calD$ was always
line-like, terminating only at material boundaries. It is worth noting
that in all cases where $\calD$ was $1$-dimensional, it was also
piecewise flat -- it is easy to argue that in order to minimize its
spatial extent, $\calD$ will in general be flat.  
Though our dimensionality arguments apply to only asymptotically thin
sheets, the predictions appear well obeyed even for sheets as thick as
$1/30$ of their width. The dimensional behavior is thus much more robust
than the ridge scaling behavior.

Because of the phantom nature of our simulated sheet, we found that
favorable energy configurations often had local geometries at vertices
in which the sheet passed through itself -- the most common example was
branched manifolds at vertices.
The arguments for scaling and the minimum dimensionality of $\calD$ 
are not affected by whether or not the sheets are phantom.
However, the set of boundary conditions that produce non-phantom 
sheets with minimal vertex dimensionality may be more limited than for
phantom $m$-sheets.

\subsection{Effect of Embedding Dimension on Scaling}

In Section~\ref{sec:scaling} we made predictions for
the energetic scaling outside of vertex structures 
for a sheet whose thickness was much smaller than its width.
However, we did not analytically address how thin the
sheet must be before it displays ``thin limit'' behavior.
Our observations of the typical progress of a numerical simulation lend
some insight into the approach to this asymptotic limit.
In the process of relaxing our sheets, we often vary the
thickness $h$, which shows us how the energy distribution
depends on $h$.  We were not able to vary $h$ enough to 
directly observe any
scaling behavior with $h$.  
Even our smallest $h$'s show little enough
of the desired asymptotic behavior, and increasing $h$ only blurs this
behavior beyond recognition.  However, the qualitative behavior with $h$
is consistent with our conclusions.  
The reported results are for the
smallest $h$ values we could reliably attain.  When $h$ is made larger,
the main effect is to reduce the dynamic range in our energy density
plots.  It makes the energy spread more uniformly over the sheet.  Where
ridge scaling is observed, increasing $h$ reduces the proportion of
stretching energy, as we've previously observed in
2-sheets\cite{Alex}.  Finally, increasing $h$ reduces the
observed effect of embedding dimension.  The clear differences between
the stretching and bending energy profiles  in 5 dimensions become
less distinct as $h$ is increased.  This is as expected.  Embedding
dimensions should have less effect if a sheet is thicker.  These
behaviors add to our confidence that the scaling behavior we report
becomes more, not less, distinct as we reduce $h$.

Our derivation of cone scaling was based on very simple and 
well-founded assumptions, 
so it was not a surprise that cone scaling was so
clearly visible in geometries where we expected to find it.
In all simulations where there was only one disclination structure in
a $3$-dimensional
manifold, the observed scaling was consistent with our predictions
based on cones. A single line disclination in either $4$ or $5$
dimensions produced a cone-structure with scaling like that of a simple
cone in a $2$-sheet. In the simulations presented in
Section~\ref{sec:singlefold}, embeddings in $5$ dimensions
produced a point-like vertex and energy scaling close to our predictions
for double-cone scaling (where there is curvature on the same order in
both material directions perpendicular to the cone generators).
The success of these predictions assures us that the cone-structure is
well understood and that, for the case of line disclinations, the
description of a relaxed configuration in terms
of a stack of $2$-sheet embeddings can be accurate.

Our simulations verified the formation of ridges in
$2$-sheets in $3$ dimensions and in 
$3$-sheets in $4$ dimensions, as witnessed in other
studies~\cite{our.stuff,eric}. 
As expected, planar ridge structures
spanning the gaps between linear vertex structures in $d=4$ had the same
energy-bearing properties as their $3$-dimensional equivalents.
However, for 
$2$-sheets in $4$ dimensions and
$3$-sheets in $5$ dimensions we consistently found no
ridge energy structures, even along what appeared to be folds.
In Fig.~\ref{fig:2sheet-embed}(c) it is apparent that the elastic 
sheet deflects
slightly into the fourth dimension in order to relieve the strain
along the ridge-center. The way this ridge decreases the
strain along its midline shows the essential difference
between embeddings of $m$ sheets in $m+1$ dimensions and in all greater
dimensions. In the geometric von Karman 
equations, Eq.~\ref{eq:GvK}, the sources of strain are
{\em sums} of curvatures along
different normals.
If there is more than one normal
direction, then there is the possibility of cancellation between
curvature terms for any given $2$-dimensional hyperplane.
On the $4$-dimensional fold the manifold
bubbles into the extra dimension, creating positive Gaussian curvature
which counters the negative Gaussian curvature of the saddle-shaped
peak-to-peak ridge profile (see for example
Fig.~\ref{fig:twosheets}). We do not assume that the cancellation of 
lowest order terms is perfect, but we showed in 
Section~\ref{sec:twosheettwodiscl}
that the stretching energy is diminished so 
greatly relative to the bending energy that ridge-like scaling is 
completely masked by conical scaling near vertices (It is notable 
that, based on the scaling arguments derived in Appendix A, in the 
limit of very large virial ratios the dominant curvature terms become 
increasingly insensitive to the elastic thickness -- much like they are
in cones). We hope to return to this subject in future research,
with the aim to find analytic expressions for
the embedding of a $2$-sheet with two disclinations in $4$ spatial
dimensions and derive the thin limit scaling directly.

\subsection{New Questions Raised by This Work}

We believe that the detailed structure of the novel point-like vertices
in $3$-sheets deserves further study. For example, in
Section~\ref{sec:singlefold} we touched on the fact that these
point-vertices are loci of folding in all three material
directions. Since the three material directions share only two normals
and are constrained by the boundary conditions that the sheet be
isometric outside the vertex point, it is likely that the angles each
direction folds by can only take {\em discrete} sets of values -- not
unlike the discretization of disclination angles in a lattice, but driven
purely by spatial geometry for an otherwise continuous medium. This
discetization would probably be more pronounced in sheets that are not
phantom, since branched manifolds at vertices add more degrees of
freedom.

Much could also be learned by viewing these new crumpling phenomena
as a mathematical boundary-layer problem, since they display several 
new and intriguing features in this light.
Our problem 
belongs to the class of variational problems given by a singularly perturbed 
energy functional $\calE^h$ with small parameter $h$.  One approach to 
analyzing such variational problems is by identifying the boundary layers 
and then determining the appropriate solutions through matched 
asymptotics. Example of this
``local'' approach, as applied to elastic $2$-sheets are the analysis
of ridges in Ref. \cite{Alex} and of the vertices in
\cite{maha.cone}.

A contrasting ``global'' approach to these problems is through the
notion of $\Gamma$ -- convergence \cite{DeG,DM}. This approach calls
for the identification of an appropriate asymptotic energy ${\cal
E}^*$ that gives the energy of a configuration in the $h \rightarrow
0$ limit. This asymptotic energy functional is called the $\Gamma$
limit of the functional ${\cal E}^h$ as $h \rightarrow 0$.
If the $\Gamma$ limit exists, the configuration of the minimizers for
a small but nonzero $h$ is then deduced by finding the minimizers for
${\cal E}^*$ and observing that the minimizer for nonzero $h$ is
close to the minimizer for ${\cal E}^*$. Note
that this approach is similar in philosophy to our arguments in
Sec.~\ref{sec:structures} where we deduced the structure of the vertex
regions for $h > 0$, by considering isometric immersions which are
relevant for $h = 0$.

Since these singularly perturbed variational problems show energy
condensation, in the limit $h \rightarrow 0$ all the energy
concentrates on to a defect set, which we denote by ${\cal
B}$. Consequently, the appropriate asymptotic energy should also be
defined for singular configurations, and it should depend on the
defect set ${\cal B}$, and the configuration $u$ outside ${\cal
B}$. Thus, the asymptotic energy is given by a functional ${\cal E}^*
= {\cal E}^{*}[h,u,{\cal B}]$. 
Examples of this type of analysis are the analysis of phase separation
in Refs.~\cite{sternberg,phase_transitions} and the asymptotic folding
energy in the context of the blistering of thin films
\cite{AG,KJ,ADM}. In both these cases, the asymptotic energy scales
with the small parameter $h$ as ${\cal E}^* \sim h^{\alpha}$ for a
fixed $\alpha$. Also, the $\Gamma$ limit is local, in that the
asymptotic energy is given by integrating a local energy density over
the defect set. 
This is in sharp contrast to the behavior of elastic
manifolds. For elastic manifolds, the asymptotic energy depends on two
kinds of defect sets, the strain defect set $\calD$ and the
curvature defect set $\calK$. The exponent $\alpha$ that gives
the scaling of ${\cal E}^*$ with $h$ is not fixed, but depends on
whether or not $\calD = \calK$. Finally, in the case of ridge
scaling, we have the following two interesting features:
\begin{enumerate}
\item The width of the boundary layer around the ridge depends on both
the small parameter $h$ and the length of the ridge $X$.
\item The energy of the ridge scales as $h^{5/3} X^{1/3}$ and is not
linear in the size of the ridge. Therefore, the asymptotic energy of
the ridge is not given by an integral of a local energy density over
$\calK$.
\end{enumerate} 
These features imply that the $\Gamma$ limit ${\cal E}^*$, if it
exists, is non-local. It is therefore very interesting to carry out a
rigorous analysis of the $\Gamma$ limit for the elastic elastic energy
functional in Eq.~\ref{eq:energy}. This analysis will probably
involve new ideas and techniques.

\section*{Conclusion}

In this paper we have found two important results for crumpled sheets. 
First, we have shown
that if the spatial dimension $d$ is greater than $m+1$, the stretching
elastic energy condenses onto vertex structures, while for the special
case $d = m+1$ it condenses onto ridges as well. 
Second, we have provided evidence that when $d < 2m$ 
the strain defect set in a crumpled sheet has dimension at least
$2m-d-1$.\tw
For higher-dimensional manifolds with $m > 3$ one may imagine
further forms of energy condensation as the embedding dimension
increases.  Such manifolds could reveal further surprises, as
the present study did.  Like gauge fields, elastic manifolds
have revealed distinctive ways in which singularities in a field
may interact at long range.  To explore further the conditions
and forms of this interaction seems worthwhile.


\acknowledgments{
The authors would like to thank Bob
Geroch, L. Mahadevan, Bob Kohn and Felix Otto for helpful and
enlightening conversations.
This work was supported in part by the National Science Foundation under 
Award number DMR-9975533 Account 5-27810.}

\appendix


\section{Derivation of Virial Relation}
\label{sec:virialderivation}

Here we derive the relation between the energy scaling exponents and the
virial ratio of bending to stretching energies on an elastic
ridge. This derivation is a generalization of the derivation presented
in~\cite{science.paper}.

We assume for simplicity that on a ridge, the elastic bending
energy is dominated by the contribution of the main curvature accross
the ridge, and this curvature is approximately constant for the
entire length of the ridge with a typical value  $C$.
For a simple ridge of length $X$ in a two dimensional manifold, 
the ridge curvature is significant in a band of width $w=1/C$ transverse
to the
ridge, so the total bending energy of the ridge is approximately
\begineq{ridge_bend_approx}
{\cal E}_b \approx \mu h^2 C^2 w X  =  \mu h^2 C X,
\end{equation}
where $\mu$ is the elastic modulus of the material and $h$ is
the thickness.

If we
assume that the ridge has a single dominant component of strain 
which also extends for a typical width $w$ and
has typical value $\gamma$, then
the total stretching energy of the ridge is
\begineq{ridge_stretch_approx}
{\cal E}_s \approx \mu \gamma^2 w X = \mu \gamma^2  X/C.
\end{equation}

We can write the
total elastic energy along the ridge as
\begin{equation}
{\cal E} \approx \mu X \left[ h^2 C + \gamma^2 C^{-1} \right] .
\end{equation}

Now, if we make a scaling ansatz, $\gamma \sim (XC)^{-\alpha}$, the 
energy becomes
\begineq{eq:app_a_2}
{\cal E} \sim \mu X \left[ h^2 C + X^{-2 \alpha} C^{-(2 \alpha +1)} \right] .
\end{equation}
We may now find the minimum energy balance 
by setting the derivative $d{\cal E}/dC$ equal to zero:
\begin{eqnarray}
\lefteqn{h^2  -(2 \alpha +1) 
X^{-2 \alpha} C^{-(2 \alpha +2)} = 0} \\
& & \Rightarrow C \sim X^{-1} \left( {h}/{X} \right)^{-1/(\alpha+1)} \\
& & \Rightarrow \gamma \sim \left( {h}/{X} \right)^{\alpha/(\alpha+1)}
\end{eqnarray}
From the first of the above equations, it is clear that the virial ratio
is related to $\alpha$ by ${\cal E}_b = (2 \alpha +1 ) {\cal E}_s$.


\begin{thebibliography}{99}
\bibitem{foot2} The broad region of the bending energy density graph
between $\approx 2 \%$ and 
$\approx 40 \%$ volume fraction which seems to show clear scaling is
well fit by a scaling exponent of $\approx -0.9$. This suggests that the
energy falls away slowly transverse to most energetic cone
generators. This observation is not contradictory to our assumption that
the region of high curvature persists for a length of order $1/C$
in the transverse directions.
\end{thebibliography}
\end{document}